\documentclass[useAMS,usenatbib,usegraphicx,onecolumn]{mn2e}
\def\simlt{\mathrel{\hbox{\rlap{\hbox{\lower5pt\hbox{$\sim$}}}\hbox{$<$}}}}
\def\simgt{\mathrel{\hbox{\rlap{\hbox{\lower5pt\hbox{$\sim$}}}\hbox{$>$}}}}
\def\be{\begin{equation}}
\def\ee{\end{equation}}

\def\BS{^{\rm BS}}
\def\H{_{\rm H}}
\def\Hubble{_{\rm Hubble}}

\def\Sch{_{\rm Sch}}
\def\Sp{_{\rm Sp}}

\def\age{_{\rm age}}
\def\b{_{\rm b}}
\def\c{_{\rm c}}
\def\coll{_{\rm coll}}
\def\colli{_{{\rm coll},i}}

\def\collmij{_{{\rm coll},>m_{ij}}}
\def\collsun{_{{\rm coll},\odot}}
\def\e{_{\rm e}}
\def\esc{_{\rm esc}}
\def\escsun{_{{\rm esc},\odot}}
\def\ij{_{ij}}
\def\multi{_{\rm multi}}
\def\obs{_{\rm obs}}
\def\peak{_{\rm peak}}
\def\pmax{^{\prime\,\max}}
\def\pr{_{\rm p}}
\def\rel{_{\rm rel}}
\def\tf{_{\rm tf}}
\def\tid{_{\rm tid}}
\def\tidsun{_{{\rm tid},\odot}}
\def\w{_{\rm w}}

\def\d{{\rm d}}

\def\km{{\rm\,km}}
\def\kms{{\rm\,km\,s^{-1}}}
\def\kpc{{\rm\,kpc}}
\def\msun{{\rm\,M_\odot}}
\def\rsun{{\rm\,R_\odot}}
\def\Lsun{{\rm\,L_\odot}}
\def\K{{\rm\,K}}
\def\mag{{\rm\,mag}}
\def\pc{{\rm\,pc}}
\def\yr{{\rm\,yr}}

\def\Gyr{{\rm\,Gyr}}

\def\bfx{{\bf x}}
\def\bfv{{\bf v}}
\def\bfr{{\bf r}}

\title{Stellar collisions in galactic centers: black hole growth and color gradients}
\author[Q. Yu]{Qingjuan Yu\thanks{Present address: Canadian Institute for
Theorectical Astrophysics, 60 St. George Street, Toronto, Ontario M5S 3H8,
Canada, yuqj@cita.utoronto.ca} \\
Princeton University Observatory, Peyton Hall, Princeton, NJ~08544-1001, USA
}

\begin{document}

\label{firstpage}
\maketitle

\begin{abstract}
\noindent
We study the effects of stellar collisions, particularly on feeding massive
black holes (BHs) and color gradients, in realistic galactic centers.
We find that the mass released by stellar collisions is not sufficient to
account for the present BH mass in galactic centers, especially in bright
galaxies.
This study, together with the study by \citet{MT99} on tidal disruption
of stars by massive BHs, implies that the material for BH growth
(especially in galaxies brighter than $\sim10^9\Lsun$) can only come from
other sources,
for example, the mass released by stellar evolution in the initial $\sim1\Gyr$
of the galaxy's lifetime, or the gas that sinks to the galactic center
in a galaxy merger.
We also analyze how the color of a stellar system is affected by collisions of
stars.
We find that collisions between main-sequence stars cannot cause observable
color gradients in the visible bands at projected radius $R\ga 0.1\arcsec$
in M31, M32 and other nearby galactic centers.
This result is consistent with the lack of an observable color gradient
in M32 at $R\ga 0.1\arcsec$.
At even smaller radii,
the color differences caused by collisions between main-sequence stars
are at most $0.08\mag$ at $R=0.02\arcsec$.
The averaged blueing due to stellar collisions
in the region $R<0.1\arcsec$ of M32 should not be larger than $0.06\mag$
in color index $U-V$ and $0.02\mag$ in $V-I$.
The observed blueing in the center of the galaxy M31
(in a $0.14\arcsec\times 0.14\arcsec$ box) must be caused
by some mechanism other than collisions between main-sequence stars.
\end{abstract}
\section{Introduction}\label{sec:intro}

\noindent 
The centers of galaxies are extreme astrophysical environments.
They house massive black holes (BHs) (e.g., \citealt{Magorrian98})
and also densely packed stars ($10^2$--$10^4\msun\pc^{-3}$ at
galactic radius $\sim 10\pc$, e.g., \citealt{Faber97}).
High stellar densities lead to frequent stellar collisions.
The outcome of the collision between two stars depends on their types
(e.g., evolutionary phase, mass and radius) and their kinematic parameters
(e.g., relative velocity and impact parameter) (e.g., \citealt{FB02}).
The two stars may both survive their collision, or the collision may 
destroy one or both of them, or lead to their coalescence.
Gas released from the colliding stars may be accreted onto the central BH,
and this process has been proposed as one of the contributions to growth of
central massive BHs (e.g., \citealt{F78}). 
Coalesced stars may appear as new stars with different
luminosity properties or different colors from their parents;
thus the luminosity and color of galactic centers are possibly affected
by stellar collisions. 
The purpose of this paper is to study the effects of stellar collisions,
particularly on feeding BHs and color gradients,
in realistic galactic centers.

Studies of central BHs in nearby galaxies have revealed a tight correlation
between central BH mass and galactic velocity dispersion
(e.g., \citealt{Tremaine02} and references therein),
which strongly suggests a close link between the formation and
evolution of galaxies and their central BHs.
In an isolated stellar system, central BHs may accrete stellar mass from
the following three sources:
(i) gas released by tidal disruption of stars and stars swallowed whole
by central BHs (mostly for stars on elongated radial orbits which may come
close to central BHs),
(ii) gas released by stellar collisions,
(iii) and gas lost by stellar winds.
Whether the above three sources are sufficient to account for the present
BH mass and which is the main contributor to the BH growth depend on 
the stellar density distribution and velocity dispersion
(e.g., \citealt{DS83,MCD91,FB02}).
The first source in this list, tidal disruption of stars,
has been studied by \citet{MT99}.
They show that the total mass of stars that are tidally disrupted or
swallowed whole over the lifetime of typical nearby galaxies is of the
order of $10^6\msun$, approximately independent of galaxy luminosity.
Thus disrupted stars may contribute significantly
to the present BH mass only in faint galaxies ($\la 10^9\Lsun$).
One part of this paper (\S~\ref{sec:BHgrowth}) studies whether the second
source (i.e., gas released by stellar collisions) is sufficient to explain
current BH masses in realistic galactic centers.

The colors of galactic centers reflect the constituents of their stellar
population, and radial color gradients may provide much information on the
formation and evolution of the stellar populations or galaxies.
In \S~\ref{sec:galcolor}, we will analyze how the color of a stellar system
is affected by collisions of main-sequence stars.
We will focus on studying two galaxies in the Local Group: M31 and M32,
which are the nearest giant spiral and dwarf elliptical galaxy
(distance: $\sim800\kpc$) and have high-resolution
observations of their centers in several color bands \citep{Lauer98}. 
Multicolor {\it HST} WFPC2 (Wide Field Planetary Camera 2) images show
that the center of M31 (the region within $\sim 1\pc$ of its central BH)
appears bluer than the rest of nucleus and the surrounding bulge.
\citet{Lauer98} suggest that this blueing is caused by collisions between
main-sequence stars since coalescence of main-sequence stars by collisions 
may form blue stragglers (which are brighter and bluer than their parent 
stars, e.g., \citealt{L89,BP95}).
However, observations show that the center of M32 lacks color gradients
at similar spatial resolution.
The lack of color gradients in M32 is inconsistent with a rough
estimate of the effects of collisions between main-sequence stars
by \citet{Lauer98},
and this inconsistency is claimed to be an important puzzle.
In \S~\ref{sec:galcolor}, we will
study whether collisions of main-sequence stars can cause observable color
gradients in the centers of M31, M32 and some other nearby galaxies.

Finally, our conclusions are given in \S~\ref{sec:discon}.

\section{Stellar collision rates and timescales}\label{sec:collrate}

\noindent
We first present a general description of stellar collision rates and
timescales.

In a stellar system, the distribution function (DF) $f(\bfx,\bfv)$ is
defined so that $f(\bfx,\bfv)\,$\- $\d^3\bfx\d^3\bfv$ is the number of stars
within a phase-space volume $\d^3\bfx\d^3\bfv$ of $(\bfx,\bfv)$. 
We define the stellar mass function of the system at time $t$ as $\Xi(m,t)$,
so that $\Xi(m,t)\,\d m$ is the probability of finding a star with mass in
the range $m\rightarrow m+\d m$ (i.e., $\int_0^\infty\Xi(m,t)\,\d m=1$);
and we define the stellar mass and radius function at time $t$ as
$\xi(m,a,t)$,
so that $\xi(m,a,t)\,\d m\d a$ is the probability of finding a star with
mass and radius in the range $(m,a)\rightarrow(m+\d m,a+\d a)$
[i.e., $\int_0^\infty\xi(m,a,t)\,\d a=\Xi(m,t)$].
We assume that the stellar mass and radius function is independent of the
location of the stars in the phase space ($\bfx,\bfv$).

The cross section for a physical collision (or contact encounter) of two
stars with mass and radius
$(m_i,a_i)$ and $(m_j,a_j)$ moving with the relative velocity at infinity
$v\rel$ is given by:
\be
\Sigma_{ij}(v\rel)=\pi b^2=\pi r\pr^2\left(1+\frac{v^2_{ij}}{v^2\rel}\right)
\simeq\cases{\pi r\pr^2  & if $v\rel\gg v\ij$ \cr
             \pi r\pr^2 v\ij^2/v\rel^2 & if $v\rel\ll v\ij$ \cr
        },
\label{eq:Sigmaij}
\ee
where $b$ is the impact parameter, $r\pr=a_i+a_j$,
and $v_{ij}=\sqrt{2G(m_i+m_j)/r\pr}$.
If $(m_i,a_i)=(m_j,a_j)$, $v_{ij}=\sqrt{2Gm_i/a_i}\equiv v\esc$ is the
escape velocity from the surface of either star.
Thus, a star $(m_j,a_j)$ moving with velocity $\bfv_j$ through a background
of stars with distribution function $f(\bfr,\bfv)$, suffers collisions
with stars with mass and radius in the range
$(m_i,a_i)\rightarrow(m_i+\d m_i,a_i+\d a_i)$ at a rate
$\Gamma_{ij}(\bfr,\bfv_j)\,\d m_i\d a_i$ given by (cf., eq.~8-116
in \citealt{BT87}):
\be
\Gamma_{ij}(\bfr,\bfv_j,t)\,\d m_i\d a_i=\xi (m_i,a_i,t)\,\d m_i\d a_i\int\d^3\bfv_i~f(\bfr,\bfv_i)|\bfv_i-\bfv_j|\Sigma_{ij}(|\bfv_i-\bfv_j|).
\label{eq:Gammaija}
\ee
If the distribution of field stars is isotropic in velocity space, we set
$v=|\bfv|$ and we have
\be
\Gamma_{ij}(\bfr,\bfv_j,t)\,\d m_i\d a_i=\xi (m_i,a_i,t)\d m_i\d a_i\int_0^\infty\d v_i~\frac{2\pi v_i}{v_j}f(\bfr,v_i)\int^{v_i+v_j}_{|v_i-v_j|}\d v\rel~v^2\rel\Sigma_{ij}(v\rel).
\label{eq:Gammaijb}
\ee
The number of collisions between two stars with masses and radii
in the ranges $(m_i,a_i)\rightarrow(m_i+\d m_i,a_i+\d a_i)$
and $(m_j,a_j)\rightarrow(m_j+\d m_j,a_j+\d a_j)$ per unit volume
per unit time, ${\cal R}_{ij}(\bfr)\,\d m_i\d a_i\d m_j\d a_j$,
is given by:
\be
{\cal R}_{ij}(\bfr,t)\,\d m_i\d a_i\d m_j\d a_j=\xi (m_j,a_j,t)\d m_i\d a_i\d m_j\d a_j\int\d^3\bfv_j~f(\bfr,\bfv_j)\Gamma_{ij}(\bfr,\bfv_j,t),
\label{eq:calRij}
\ee
and we have ${\cal R}_{ij}={\cal R}_{ji}$.
We define the collision timescale of the stars with mass and radius $(m_i,a_i)$
as follows:
\be
t\colli(\bfr,t)\equiv\frac{\xi (m_i,a_i,t)n(\bfr,t)}{\int \d m_j\d a_j~{\cal R}_{ji}(\bfr,t)},
\label{eq:tcolli}
\ee
where $n(\bfr,t)\equiv \int\d^3\bfv~f(\bfr,\bfv)$ is the stellar
number density at position $\bfr$ and time $t$.
Thus the total number density of the stars with mass and radius in the range
$(m_i,a_i)\rightarrow(m_i+\d m_i,a_i+\d a_i)$ per unit time undergoing
collisions is given by
$n(\bfr,t)\xi(m_i,a_i,t)\d m_i\d a_i/t\coll{_{,i}}(\bfr,t)$.
For the whole stellar population, the collision timescale $t\coll(\bfr)$ 
is defined by:
\be
t\coll(\bfr,t)\equiv\frac{n(\bfr,t)}
{\int\d m_i\d a_i\d m_j\d a_j~{\cal R}_{ij}(\bfr,t)}
=\frac{1}{\int\d m_i\d a_i~\xi(m_i,a_i,t)/t\colli(\bfr, t)},
\label{eq:tcoll}
\ee
and the total number of stars per unit time undergoing collisions
is given by:
\be
\dot N\coll(t)=\int\d^3\bfr~\frac{n(\bfr,t)}{t\coll(\bfr,t)}.
\label{eq:dotNcoll}
\ee

Note that stellar collision rates depend on the position $\bfr$ in the stellar
system and hence vary during the stellar orbit (unless the orbit is circular).

\section{Feeding central massive BHs by stellar collisions}
\label{sec:BHgrowth}

\noindent
In this section, we will study whether the mass released by stellar
collisions makes a significant contribution to the growth of massive BHs
in galactic centers.

\subsection{Stellar mass involved in collisions}\label{sec:masscoll}

\noindent
The mass of gas released by two colliding stars depends on their relative
velocity and impact parameter, as well as their masses and radii and their
evolutionary phases (e.g., main-sequence stage or post-main-sequence stage)
(e.g., \citealt{MCD91,BD99,FB02}).
We define $\Delta m_{ij}(\bfr)$ as the average gas mass released per
collision between two stars with masses and radii $(m_i,a_i)$ and $(m_j,a_j)$
at position $\bfr$.
The total gas mass released by stellar collisions per unit volume
per unit time is given by:
\be
\dot\rho\coll(\bfr,t)={1 \over 2}\int \d m_i\d a_i \d m_j\d a_j~{\cal R}_{ij}(\bfr,t)\Delta m_{ij}(\bfr),
\label{eq:dotrhocoll}
\ee
where the collision rate ${\cal R}_{ij}(\bfr,t)$ can be obtained from equation
(\ref{eq:calRij}) and we have the factor ``1/2'' in front of the integration
because the gas released from each colliding star is counted twice
in the integration.
The total collisionally released mass per unit volume at
position $\bfr$ until time $t$ is given by
$\rho\coll(\bfr,t)=\int_0^t \dot\rho\coll(\bfr,t')~\d t'$,
and the total collisionally released mass over the age of the stellar
system $T\age$ is given by:
\be
M\coll(T\age)=\int \d^3\bfr~\rho\coll(\bfr,T\age).
\label{eq:Mcoll}
\ee
If all the released gas is accreted onto the central BH, equation
(\ref{eq:Mcoll}) gives the mass growth of the central BH caused
by stellar collisions.
If $\Delta m_{ij}=m_i+m_j$, equation (\ref{eq:Mcoll}) gives the
stellar mass involved or maximum gas mass released in collisions,
$M\coll^{\max}(T\age)$.

To estimate an upper limit on the BH mass growth caused by stellar
collisions, we will assume below that all the mass involved in collisions
is released as gas and accreted onto the central BH.
According to equation (\ref{eq:tcoll}),
the variation of the stellar number density is given by
\be
\dot n(\bfr,t)=-n(\bfr,t)/t\coll(\bfr,t).
\label{eq:dotn}
\ee
In equation (\ref{eq:dotn}), we have assumed that the stellar density
at a position $\bfr$ is affected only by collisions occurring near $\bfr$,
and we will see in \S~\ref{sec:resBHgrowth} that our conclusions will not
be significantly affected after relaxing this assumption.
Now, we assume that the stars in the stellar system have a single mass $m_*$
and a single radius $a_*$ [i.e., $\xi(m,a,t)=\delta(m-m_*)\delta(a-a_*)$,
where $\delta(x)$ is the Dirac function] and
the stellar system initially has an isothermal distribution
$f(\bfr,\bfv)=n(\bfr,0)\exp(-|\bfv|^2/2\sigma^2)/(2\pi\sigma^2)^{3/2}$
(where $\sigma$ is the one-dimensional velocity dispersion).
The initial collision timescale in this stellar system, which can be derived
from equations (\ref{eq:Sigmaij})--(\ref{eq:tcoll}), is given by
(see also eq.~20 in \citealt{DS83}):
\be
t\coll(\bfr,0)=\frac{1}{16\sqrt{\pi} n(\bfr,0)a_*^2 \sigma[1+v\esc^2/(4\sigma^2)]}.
\label{eq:tcolliso}
\ee
In such a stellar system, if we ignore star formation, stellar evolution,
gradual mass segregation or other evolutionary effects, then
according to equation (\ref{eq:tcoll})
[i.e., $t\coll(\bfr,t)\propto 1/n(\bfr,t)$] and equation (\ref{eq:dotn}),
we find the stellar number density at time $t$ given by:
\be
n(\bfr,t)=\frac{n(\bfr,0)}{1+t/t\coll(\bfr,0)},
\label{eq:nrt}
\ee
and the stellar collision timescale is given by
$t\coll(\bfr,t)=t+t\coll(\bfr,0)$.
Thus, in the region where the initial collision timescale is much longer
than the age of the stellar system [i.e., $t\coll(\bfr,0)\gg T\age$],
we have $n(\bfr,T\age)\simeq n(\bfr,0)$
(i.e., the stellar number density is affected little by collisions)
and the collision timescale $t\coll(\bfr,T\age)\simeq t\coll(\bfr,0)$;
while, in the region with $t\coll(\bfr,0)\ll T\age$, we have
$n(\bfr,T\age)\simeq n(\bfr,0)t\coll(\bfr,0)/T\age\ll n(\bfr,0)$ 
(i.e., stars are disrupted quickly by stellar collisions)
and the collision timescale is $t\coll(\bfr,T\age)\simeq T\age$.
The total gas mass released by collisions (or the upper limit of the total
stellar mass involved in collisions) per unit volume over the age of
the stellar system is given by:
\begin{eqnarray}
\rho\coll^{\max}(\bfr,t=T\age) & = & m_*[n(\bfr,0)-n(\bfr,T\age)] \nonumber \\
& \simeq &
\cases{ m_*n(\bfr,0)T\age/t\coll(\bfr,0) & if $t\coll(\bfr,0)\gg T\age$ \cr
        m_*n(\bfr,0)  & if $t\coll(\bfr,0)\ll T\age$ \cr
        },
\label{eq:rhocoll}
\end{eqnarray}
and the total mass involved in stellar collisions over time $T\age$ is
given by:
\be
M\coll^{\max}(T\age)=\int \d\bfr^3~\rho\coll^{\max}(\bfr,T\age)
\label{eq:Mcollmax}
\ee
If the initial stellar density is singular, that is, 
\be
n(r,0)=\frac{\sigma^2}{2\pi Gm_*r^2}
\label{eq:nsig}
\ee
(cf., eq.~4-123 in \citealt{BT87}),
then using equations (\ref{eq:tcolliso})--(\ref{eq:nsig}), we have
the total mass involved in stellar collisions over time $T\age$ as follows:
\be
M\coll^{\max}(T\age)=\sqrt{A}\sigma^2/G,
\label{eq:Mcolliso}
\ee
where
\be
A=\frac{8\pi^{3/2}\sigma^3a_*^2T\age}{Gm_*}\left(1+\frac{v\esc^2}{4\sigma^2}\right).
\label{eq:A}
\ee

In a stellar system with a distribution of stellar masses and radii,
if we ignore star formation, stellar evolution, gradual mass segregation or
other evolutionary effects and
we assume that collision timescales of stars with any given mass and radius
$(m_i,a_i)$ do not change with time,
i.e., $t\colli(r,t)=t\colli(r,0)$,
we may obtain the number density of stars with mass and radius $(m_i,a_i)$
at time $t$ is $\Xi(m_i,0) n(r,0){\rm e}^{-t/t\colli(r,0)}$
(cf., eq.~\ref{eq:dotn}),
and the stellar mass involved in collisions over the age of the
stellar system is given as follows:
\be
M\coll^{\max}(T\age)=\int 4\pi r^2\d r\int\d m_i~m_i
\Xi(m_i,0) n(r,0) \left[1-{\rm e}^{-T\age/t\colli(r,0)}
\right].
\label{eq:Mcolldis}
\ee
In reality, the collision timescale $t\colli(r,t)$ is
not a constant and should increase with time,
$M\coll^{\max}(T\age)$ obtained from equation (\ref{eq:Mcolldis})
is an upper limit to the stellar mass involved in collisions.

Ignoring star formation and stellar evolution in the study is usually
a reasonable assumption for an old stellar system after the initial rapid
stellar evolution of $\sim 1\Gyr$ is complete (i.e., massive stars have
evolved into stellar remnants and the main-sequence stars have low masses
with long lifetime and little change in the mass and radius in the
main-sequence phase; cf., \S~\ref{subsec:gendis}).
Stellar remnants of massive stars (BHs and neutron stars) may migrate
inwards as a result of energy equipartition with low-mass
field stars (e.g., \citealt{M93}), i.e., mass segregation. 
As will be discussed at the end of \S~\ref{subsec:gendis},
mass segregation affects little on our result of $M\coll^{\max}$
in most galactic centers, especially for bright galaxies.

\subsection{Galaxy samples} \label{sec:galsmp}

\noindent
In the past several years, images from the {\it Hubble Space Telescope}
({\it HST}) have revealed many details about the central regions of nearby
galaxies, with a resolution of 0.1\arcsec, corresponding to distances of
$\la 10\pc$ or $\sim10^6(M_\bullet/10^8\msun)$ Schwarzschild radii for
typical target galaxies \citep{Byun96}.
The inner surface brightness profiles of the galaxies as a function of
projected radius $R$ are well fitted with a
five-parameter fitting function --- the Nuker law \citep{Faber97}:
\be
I(R)=2^{\frac{\beta-\gamma}{\alpha}}I\b\left(\frac{R}{r\b}\right)^{-\gamma}
\left[1+\left(\frac{R}{r\b}\right)^\alpha\right]^{-\frac{\beta-\gamma}{\alpha}}.
\label{eq:nukerlaw}
\ee
The asymptotic logarithmic slope inside $r\b$ is $-\gamma$, the asymptotic
outer slope is $-\beta$, and the parameter $\alpha$ characterizes
the sharpness of the break. The break radius $r\b$ is the point of maximum
curvature in log-log coordinates. The ``break surface brightness'' $I\b$ is the
surface brightness at $r\b$. 
According to the inner surface brightness profiles $I(R)\propto R^{-\gamma}$
($r\rightarrow 0$), elliptical and spiral bulges (hot galaxies) can be
classified into two types: core galaxies ($\gamma\la0.3$) and power-law
galaxies ($\gamma\ga0.5$).
A sample of nearby hot galaxies (elliptical galaxies or spiral
bulges) observed by {\it HST} is collected in Faber et al. (1997),
who estimate the stellar mass-to-light ratio $\Upsilon$
(constant for each galaxy, determined by
normalizing to the central velocity dispersion based on spherical and
isotropic models fitted to the Nuker-law profile) and the half-light
radius (cf., Table 1 in \citealt{Y02}).
Assuming that the galaxies are spherical and isotropic
(the mean ellipticity is 0.26), we may obtain their stellar
density and velocity distributions by using the Eddington formula
(eq.~4-140a in \citealt{BT87})
and estimate their central BH masses $M_\bullet$
by using the following relation between BH mass and galactic velocity
dispersion \citep{Tremaine02}
\be
\log (M_\bullet/\msun)=(8.13\pm 0.06)+(4.02\pm 0.32)\log (\sigma\e/200\kms),
\label{eq:msigma}
\ee
where $\sigma\e$ is the luminosity-weighted line-of-sight velocity dispersion
inside the half-light radius.
Note that the surface brightness profiles obtained in \citet{Faber97} are
based on {\it HST} WFPC1 data.
We have updated the surface brightness profile of M32 using
WFPC2 data in this paper (for detailed information,
see \S~\ref{subsec:M32} or \citealt{Lauer98}). 
In total, 31 galaxies from the Faber et al.\ sample will be used in this
paper to study the effects of stellar collisions in galactic centers.

\subsection{Results for simple stellar systems with identical stars}
\label{sec:resBHgrowth}

\begin{figure}
\begin{center}
\includegraphics[width=0.7\textwidth,angle=0]{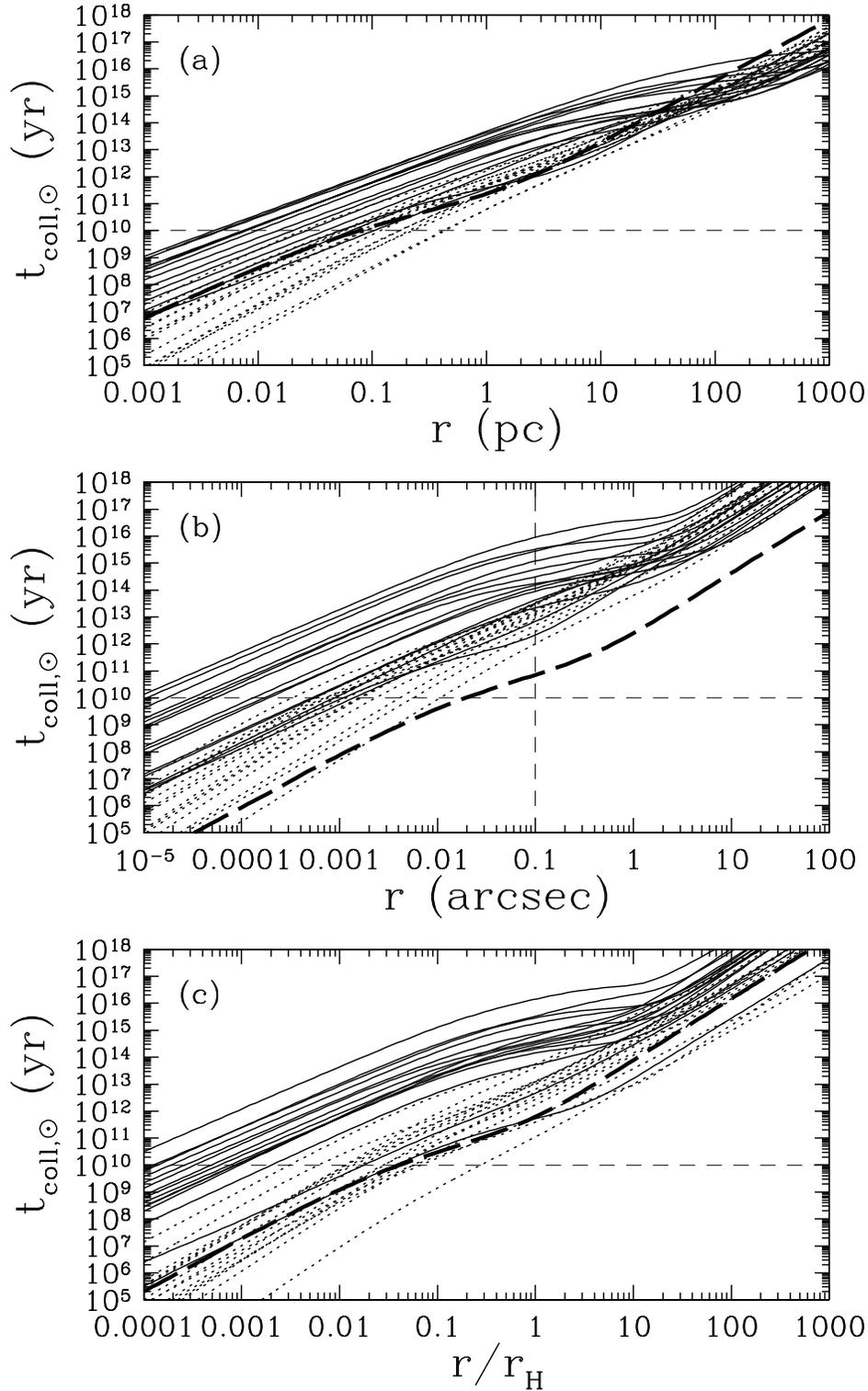}
\caption{Stellar collision timescales as a function of galactic radius
(eq.~\ref{eq:tcoll}).
The galaxy sample comes from Table 1 in Yu (2002).
Stars are assumed to all have the solar mass and radius.
In the region not resolved by the {\it HST} ($2r<0.1\arcsec$),
the collision timescales are obtained by extrapolating the surface brightness
profiles (eq.~\ref{eq:nukerlaw}) inward.
The solid curves represent core galaxies ($\gamma\simlt 0.3$) and
the dotted curves represent power-law galaxies ($\gamma\simgt 0.5$).
The thick dashed line represents the galaxy M32, whose central colors will
be studied in detail in \S~\ref{subsec:M32}.
The three panels use three different measures of radius: parsecs, arcsec
and $r\H$, where $r\H$ is the radius of the sphere of influence of the BH,
defined implicitly in terms of the intrinsic one-dimensional velocity
dispersion of the galaxy $\sigma(r)$ through equation (\ref{eq:sigmaH}).
For how the results shown in this Figure are affected after generalizing
to a distribution of masses and radii, see \S~\ref{subsec:gendis}.
}
\label{fig:tcoll}
\end{center}
\end{figure}

\noindent
In this subsection, we use equation (\ref{eq:tcoll})
to find the collision timescales as a function of
galactic radius in realistic galactic centers (cf., \S~\ref{sec:galsmp}).
The results are shown in Figure~\ref{fig:tcoll}.
Then, we will use equations (\ref{eq:dotrhocoll})--(\ref{eq:A}) to obtain
the stellar mass involved in collisions or the upper limit of mass released by
collisions ($M\coll^{\max}$) over the Hubble time $t\Hubble\simeq10^{10}\yr$,
which is shown in Figure~\ref{fig:Mcoll}.
In this subsection, all the stars are assumed to have
identical mass and radius and stellar evolution is ignored.
In the next subsection, we will see that the generalization to a
distribution of masses and radii will not significantly affect our
conclusions.

Figure~\ref{fig:tcoll} shows the collision timescales $t\collsun(r)$
as a function of radius $r$
(the subscript ``$\odot$'' indicates that the results are obtained by assuming
that all stars have the solar mass and radius).
The stellar distribution $n(r)$ and the collision timescale
$t\collsun(r)$ within the region not resolved by the
{\it HST} ($2r<0.1$\arcsec) are obtained by extrapolating the observed surface
brightness profile (eq.~\ref{eq:nukerlaw}) inward
[i.e., assuming $I(R)\propto R^{-\gamma}$ at $2R<0.1$\arcsec].
As seen from Figure~\ref{fig:tcoll}, the collision timescales $t\collsun(r)$
increase with increasing radius $r$.
At radii $r\ga 0.003$--$1\pc$ (Fig.~\ref{fig:tcoll}a),
the collision timescales are longer
than the Hubble time and the stellar distributions are affected little
by stellar collisions [i.e., $n(\bfr,0)\simeq n(\bfr,t\Hubble)$ and
$t\coll(\bfr,0)\simeq t\coll(\bfr,t\Hubble)$],
which is also true in the region resolved by {\it HST}
($2r>0.1\arcsec$, cf., Fig.~\ref{fig:tcoll}b).
Figure~\ref{fig:tcoll}(c) shows that in all galaxies, the region with
$t\collsun(r)<t\Hubble$ is located within the radius of the sphere of
influence of the BH $r\H$,
which is defined in terms of the intrinsic one-dimensional velocity dispersions
of the galaxy $\sigma(r)$ through
\be
\sigma^2(r\H)\equiv GM_\bullet/r\H.
\label{eq:sigmaH}
\ee
In the region where most stars should have been disrupted quickly
[i.e., $t\collsun(r)\ll t\Hubble$],
the stellar distribution obtained by extrapolating the Nuker law inward
can only be interpreted as the initial stellar distribution and the
corresponding collision timescales $t\collsun(r)$ shown in
Figure~\ref{fig:tcoll} are the initial collision timescales
$t\collsun(r,0)$.
Current stellar densities in those regions may be estimated from equation
(\ref{eq:nrt}), and current collision timescales are about
$t\coll(r,t\Hubble)\simeq t\Hubble$.

\begin{figure}
\begin{center}
\includegraphics[width=0.8\textwidth,angle=0]{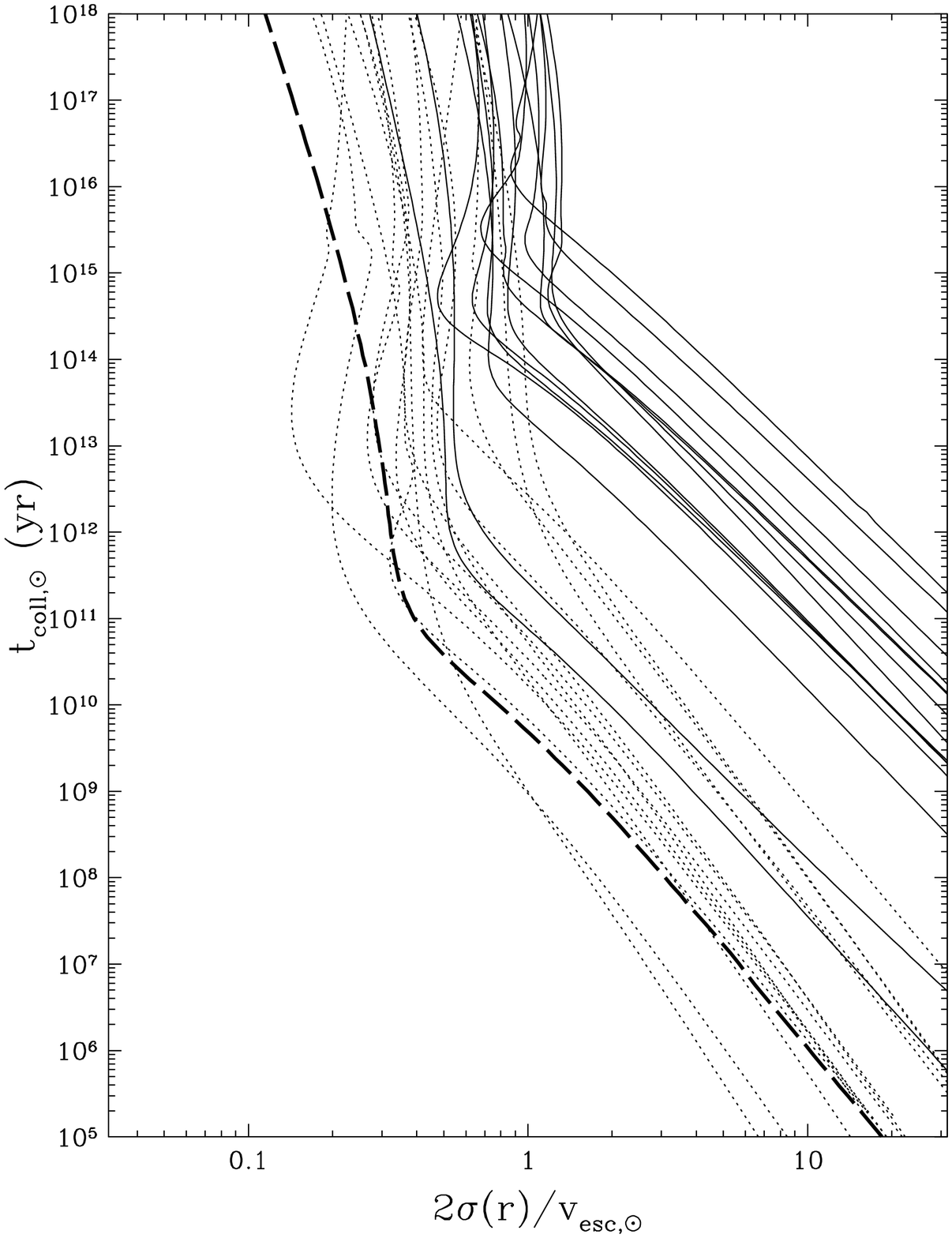}
\caption{Stellar collision timescales $t\collsun$ as a function of 
the ratio $\sigma(r)/v\escsun$.
The quantity $\sigma(r)$ is the one-dimensional velocity dispersion
and $v\escsun\simeq 620\kms$ is the escape velocity from stars with solar mass
and radius (cf., eq.~\ref{eq:Sigmaij}).
The galaxies are the same as those in Figure~\ref{fig:tcoll}.
Stars are assumed to all have the solar mass and radius.
The line types have the same meaning as those in Figure~\ref{fig:tcoll}.
This figure shows that the velocity dispersion $\sigma(r)$
is not a monotonic function of radius $r$,
which is due to the effect of central BHs
(note that $t\collsun$ is a monotonic function of $r$ in
Fig.~\ref{fig:tcoll}).
}
\label{fig:tcollsigma}
\end{center}
\end{figure}

Figure~\ref{fig:tcollsigma} shows the collision timescale
$t\collsun$ as a function of the ratio $2\sigma(r)/v\escsun$,
where $v\escsun$ is the escape velocity from stars with solar mass and radius
(cf., eq.~\ref{eq:Sigmaij}).
As seen from Figure~\ref{fig:tcollsigma}, the one-dimensional
velocity dispersion $\sigma(r)$ is not a monotonic function of
$t\collsun(r)$ or $r$ (note that $t\collsun$ is a monotonic increasing
function of $r$ in Fig.~\ref{fig:tcoll}),
which is due to the effect of central BHs.
In the absence of central BHs, the velocity dispersion $\sigma(r)$ 
should decrease with decreasing radius $r$ at small $r$
(cf., eq.~10 for case $1<\eta<2$ in \citealt{Tremaine94});
but with a central BH,
the velocity dispersion increases with decreasing $r$
at $r\ll r\H$ [i.e. $\sigma(r)\propto r^{-1/2}$].
Thus, with decreasing $r$ or $t\collsun(r)$,
the velocity dispersion $\sigma(r)$ may show a minimum near
the region where the effect of central BHs becomes dominant.
Figure~\ref{fig:tcollsigma} shows that for all the core galaxies,
the collision time is less than the Hubble time
only if the velocity dispersion exceeds the escape speed.

\begin{figure}
\begin{center}
\includegraphics[width=0.8\textwidth,angle=0]{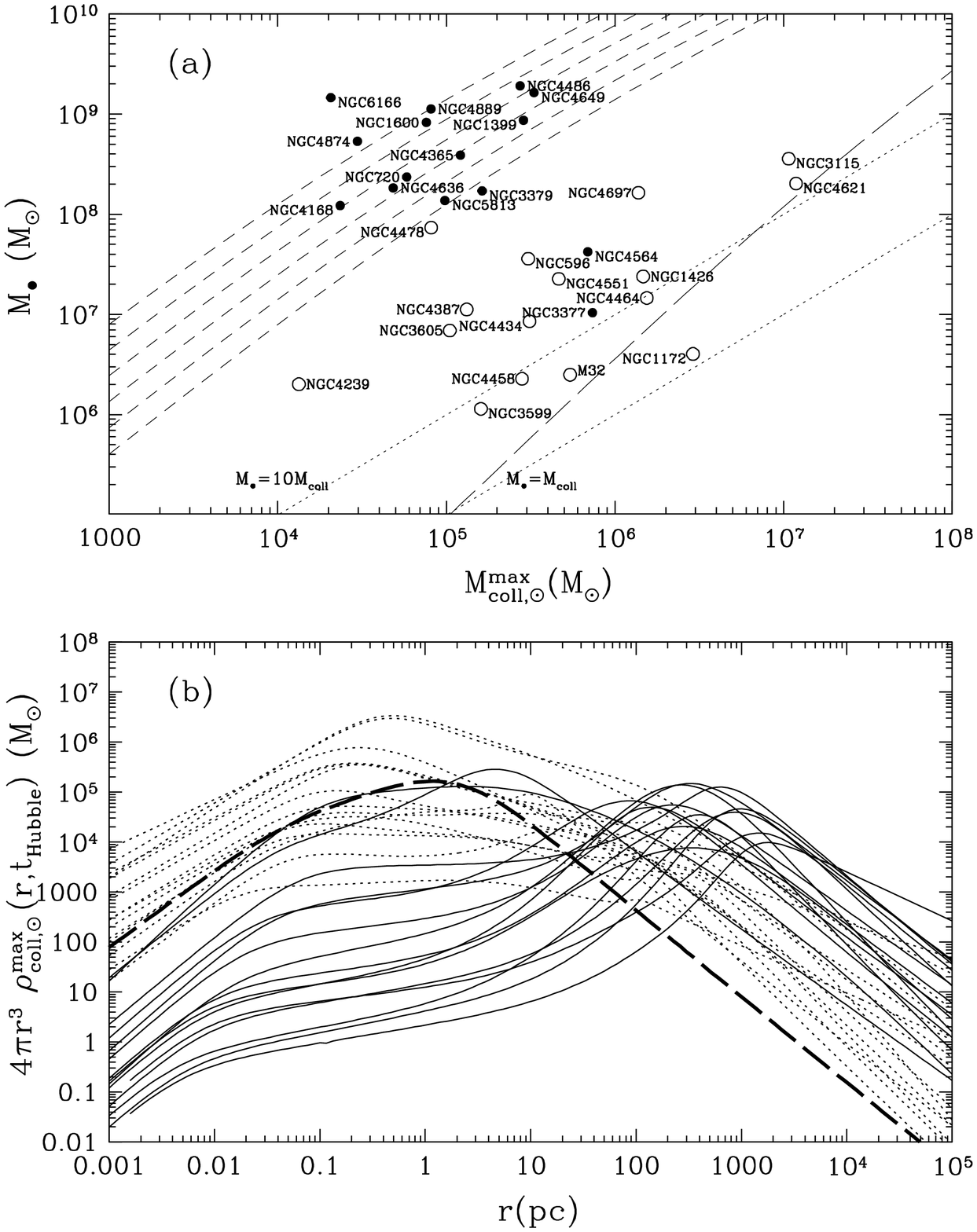}
\caption{Panel (a) shows the total stellar mass involved in collisions
$M\collsun^{\max}(t\Hubble)$ (eq.~\ref{eq:Mcoll}) over a Hubble time,
versus the central BH mass $M_\bullet$.
The galaxies are the same as those in Figure~\ref{fig:tcoll}.
The BH mass $M_\bullet$ is obtained by the BH mass versus galactic velocity
dispersion relation (eq.~\ref{eq:msigma}) in \citet{Tremaine02}. 
Stars are assumed to have the solar mass and radius.
The stellar mass involved in collisions is an upper limit to the mass
that collisions can contribute to the BH growth
(assuming both stars are completely disrupted and the entire gas mass is
accreted by the BH).
The solid circles represent core galaxies; and the open circles represent
power-law galaxies.
The dotted lines are the reference lines representing
$M\collsun^{\max}(t\Hubble)=M_\bullet$ and
$M\collsun^{\max}(t\Hubble)=0.1M_\bullet$.
The dashed curves give the relation between $M_\bullet$ and
$M\collsun^{\max}(t\Hubble)$ in simplified galaxy models with isothermal
stellar distributions.
The long dashed curve gives the results with singular stellar densities
(see eqs.~\ref{eq:nsig}--\ref{eq:A}).
The short dashed curves give the results with flat inner stellar densities
(eq.~\ref{eq:nflat});
reading upwards, the core radius $r\c$ in equation (\ref{eq:nflat})
varies from $10^2\pc$ to $10^3\pc$
with interval $\Delta\log (r\c/\pc)=0.2$.
In all cases $M\collsun^{\max}(t\Hubble)<M_\bullet$.
Panel (b) shows $4\pi r^3\rho^{\max}\collsun(r,t\Hubble)$ as a function of
radius $r$ (cf., eqs.~\ref{eq:dotrhocoll} and \ref{eq:Mcoll}).
The peaks of the curves give the location where most of the stellar mass
involved in collisions originates.
The line types have the same meaning as those in Figure~\ref{fig:tcoll}.
For how the results shown in this Figure change after generalizing
to a distribution of masses and radii, see \S~\ref{subsec:gendis}.
}
\label{fig:Mcoll}
\end{center}
\end{figure}

Figure~\ref{fig:Mcoll}(a) shows the central BH mass $M_\bullet$ versus
the upper limit of the mass released by stellar collisions over
a Hubble time $M\collsun^{\max}(t\Hubble)$ (eq.~\ref{eq:Mcoll})
in the galaxies in our sample (open circles and solid circles).
Once again, all stars are assumed to have the solar mass and radius.
As seen from Figure~\ref{fig:Mcoll}(a),
the upper limit of the mass released by stellar collisions
over a Hubble time $M\collsun^{\max}(t\Hubble)$ is
smaller than the BH mass $M_\bullet$ in {\em all} the galaxies of the sample.
For power-law galaxies ($M_\bullet\sim10^5$--$10^9\msun$, open circles),
$M\collsun^{\max}$ is usually in the range $\sim10^4$--$10^7\msun$;
and for core galaxies ($M_\bullet\sim10^8$--$10^9\msun$, solid circles),
$M\collsun^{\max}\sim10^4$--$10^6\msun$.
Thus, other material sources, rather than mass released by stellar collisions,
must dominate the growth of central BHs, especially in core galaxies.
In Figure~\ref{fig:Mcoll}(a), we also show the results obtained from
simplified galaxy models with Maxwellian stellar distributions with
dispersion independent of radius (dashed lines).
The short dashed lines are obtained by assuming that the galaxies have
flat inner stellar densities 
(see eqs.~4-124b and 4-128a in \citealt{BT87})
\be
n(r)=n\c/[1+(r/r\c)^2]^{3/2},
\label{eq:nflat}
\ee
where $n\c=9\sigma^2/(4\pi G m_* r\c^2)$.
The core radius $r\c$ is set to be in the range $10^2$--$10^3\pc$.
Note that it is only in the singular case
that the density in equation (\ref{eq:nsig})
and a Maxwellian velocity distribution are an exact solution
of the collisionless Boltzmann equation.
The long dashed line is obtained by assuming that the galaxies have
singular stellar densities (see eqs.~\ref{eq:nsig}--\ref{eq:A}).
Here, the isothermal distributions with the flat and singular stellar densities
represent simplified cases for core and power-law galaxies, respectively.
As seen from Figure~\ref{fig:Mcoll}(a), most of the solid circles
(obtained from actual core galaxies) are located in the region
covered by the short dashed lines
(those solid circles not covered have break radii $r\b$
out of the assumed range of $r\c\sim 10^2$--$10^3\pc$).
The mass involved in collisions
obtained from the singular isothermal distribution (long dashed line),
is basically an upper bound to the masses obtained from actual power-law
galaxies (open circles),
because the singular stellar density profile is generally steeper
than the inner stellar density profiles of most power-law galaxies.

Figure~\ref{fig:Mcoll}(b) shows $4\pi r^3\rho\collsun^{\max}(r, t\Hubble)$
(cf., eqs.~\ref{eq:dotrhocoll} and \ref{eq:Mcoll}) as a function of galactic
radius $r$.
The peaks or highest parts of these curves represent the regions where most
of the stellar mass involved in collisions comes from.
The radii where the peaks of the curves are located can be denoted by $r\peak$.
We have $\langle v\rel^2\rangle=6\sigma^2(r)$ for the typical relative
velocity of two colliding stars at radius $r$.
As seen from Figure~\ref{fig:Mcoll}(b),
for core galaxies (solid lines), the peaks of $4\pi r^3\rho\collsun^{\max}(r)$
are generally located at large radii ($r\peak\sim10^2$--$10^3\pc$).
For power-law galaxies (dotted lines), some of the curves show apparent peaks
at small radii ($r\peak\sim0.1$--$1\pc$);
while some have flat-topped profiles covering a
large range of radii (e.g., $10^{-2}$--$10^2\pc$).
Note that the mass-to-light ratios of the galaxies are obtained by
normalizing to the central velocity dispersion based on spherical and
isotropic models fitted to the Nuker-law surface brightness profiles.
Thus, given the surface brightness profile and the mass-to-light ratio of
the stellar system, we have
at $r\ll r\b$ ($r\b$: the break radius, cf., eq.~\ref{eq:nukerlaw})
$n(r)\propto r^{-1}I(r)/m_*\propto r^{-\gamma-1}/m_*$ for power-law galaxies.
Now, consider two regions with $r\ll r\b$ in power-law galaxies: 
\begin{itemize}
\item In a region with $t\coll(r)\ll t\Hubble$, we have
  \be
  r^3\rho\coll^{\max}(r)\propto r^3n(r)m_*\propto r^{-\gamma+2},
  \label{eq:r3rhocolla}
  \ee
  and since $\gamma<2$, $r^3\rho\coll^{\max}(r)$ increases with increasing
  galactic radii $r$ (this applies to the region $r\la0.01$--$0.1\pc$
  in Figure~\ref{fig:Mcoll}b).
\item In a region with $t\coll(r)\gg t\Hubble$, we have
(from eqs.~\ref{eq:Sigmaij}, \ref{eq:tcoll} and \ref{eq:nukerlaw})
  \be
  t\coll(r)\propto [n(r)\sigma(r)\Sigma_{ij}]^{-1} \propto
  \cases{ r^{\gamma+1}\sigma(r)^{-1}m_*/a_*^2 & if $\sigma(r)\gg v\esc$ \cr
          r^{\gamma+1}\sigma(r)/a_*  & if $\sigma(r)\ll v\esc$, \cr 
        }
  \label{eq:tcollmstar}
  \ee
  and
  \begin{eqnarray}
  r^3\rho\coll^{\max}(r) & \propto & r^3n(r)m_*/t\coll(r) \propto
  \cases{ 
   r^{-2\gamma+1}\sigma(r)a_*^2/m_* & if $\sigma(r)\gg v\esc$ \cr
   r^{-2\gamma+1}\sigma(r)^{-1}a_* & if $\sigma(r)\ll v\esc$, \cr
        }
  \label{eq:r3rhocollb}
  \end{eqnarray}
  The (one-dimensional) velocity dispersion $\sigma(r)$ varies only slowly
  with the galactic radius $r$ at $r>r\H$, and varies as $r^{-1/2}$
  at $r<r\H$, where $r\H$ is the radius of the sphere of influence of the BH
  (cf., eq.~\ref{eq:sigmaH}).
  In the region with $\sigma(r)\gg v\esc(r)$, we usually have $r\ll r\H$.
  Thus, according to equation (\ref{eq:r3rhocollb}), with increasing $r$,
  $r^3\rho\coll^{\max}(r)$ monotonically
  decreases with radius for power-law galaxies with $\gamma\ga 0.75$.
  For power-law galaxies with $0.75\ga\gamma\ga0.5$,
  $r^3\rho\coll^{\max}(r)$ is usually not a monotonic increasing or decreasing
  function of $r$ due to the different variation of $\sigma(r)$ with $r$
  at $r<r\H$ and $r>r\H$.
\end{itemize}
For power-law galaxies with $\gamma\ga 0.75$, since $r^3\rho\coll^{\max}(r)$
is an increasing function of $r$ in the region with $t\coll(r)\ll t\Hubble$
and is a decreasing function of $r$ in the region $r<r\b$ with
$t\coll(r)\gg t\Hubble$,
$r^3\rho\coll^{\max}(r)$ shows a peak in the region with
$t\coll(r)\simeq t\Hubble$,
and the mass involved in collisions comes mainly from small radii
$r\peak\sim$0.1--1$\pc$ (cf., Fig.~\ref{fig:tcoll}a)
(see Figure~\ref{fig:Mcoll}b).
However, for core galaxies, the peak of $r^3\rho\collsun^{\max}(r)$
is generally not located around the radii with $t\coll(r)\simeq t\Hubble$,
but at large radii $r\peak\sim10^2$--$10^3\pc$ where the stellar densities
show a break (see the break radii $r\b$ in eq.~\ref{eq:nukerlaw}).
In our analysis below, we will focus on power-law galaxies with
$\gamma\ga 0.75$ and core galaxies, which usually can be analytically
studied; the results for power-law galaxies with $0.75\ga \gamma\ga 0.5$
are usually intermediate cases between the results of those two types of
galaxies.

Note that in equation (\ref{eq:dotn}), we assume that the stellar density
$n(\bfr)$ is affected only by local stellar collisions.
This assumption will not underestimate the mass involved in collisions
per unit volume at position $\bfr$
if the collision timescale at position $\bfr$
is much longer than the age of the stellar system,
and this assumption will not affect the estimate of the mass involved
in collisions per unit volume at position $\bfr$
if stars are on circular orbits.
For core galaxies, the mass involved in collisions $M\coll^{\max}$ comes
mainly from near the break radii $r\b$,
where $t\coll(r\b,0)\gg t\Hubble$;
hence, $M\coll^{\max}\simeq m_*n(r\b)t\Hubble/t\coll(r\b,0)$ will not
be affected even if we relax the assumption in equation (\ref{eq:dotn}).
For power-law galaxies with $\gamma\ga0.75$,
the mass involved in collisions $M\coll^{\max}$ comes mainly from the
radii $r\peak$ where $t\coll(r\peak)\simeq t\Hubble$,
and $M\coll^{\max}$ is approximately equal to the initial stellar mass
within the radius $r\peak$ (cf., eqs.~\ref{eq:rhocoll} and \ref{eq:Mcollmax}).
Note that the stars on eccentric orbits which are not involved in
collisions at their apocenters (e.g., $>r\peak$),
may be involved in collisions at their pericenters (e.g., $<r\peak$), where
the collision timescale is shorter due to high stellar density;
and the mass involved in collisions in Figure~\ref{fig:Mcoll}(a)
is therefore possibly an underestimate.
However, the total stellar mass involved in collisions at radius $r$ cannot
exceed the total mass of stars which may come within radius $r$.
To find an upper limit to the correction needed for non-locality,
we assume that {\em all} the stars are involved in collisions at their
pericenters
(if the collision timescale at their pericenter is $t\coll(r)<t\Hubble$)
and the total mass involved in collisions per unit
volume at position $\bfr$ is about
$\min[m_*n(r,0)t\Hubble/t\coll(r,0), (1/4\pi r^2)\d M_{<r}(r)/\d r]$,
where $M_{<r}(r)$ represents the total mass of stars which may come within
radius $r$ in spherical stellar systems (cf., Fig.~1a in \citealt{Y02}).
We find that with these assumptions, $M\coll^{\max}$ for power-law galaxies
with $\gamma\ga 0.75$ in Figure~\ref{fig:Mcoll}(a) will increase at most
by a factor of 2--5 (average 3).

The collision timescale and the stellar mass involved in collisions
shown in Figures~\ref{fig:tcoll} and \ref{fig:Mcoll} are obtained by
assuming that all of the stars have the solar mass and radius.
Now consider the more general case in which all the stars have identical
mass and radius ($m_*,a_*$), which are not necessarily the solar values.
We find the following changes in the results shown
in Figures \ref{fig:tcoll} and \ref{fig:Mcoll}:

\begin{itemize}
\item The collision timescales vary as
$(m_*/\msun)(\rsun/a_*)^2$ [if $v\esc\ll\sigma(r)$] or $\rsun/a_*$
[if $v\esc\gg\sigma(r)$] times $t\collsun(r)$ shown in
Figure~\ref{fig:tcoll} (cf., eq.~\ref{eq:tcollmstar}).

\item In the region with $t\coll(r)\ll t\Hubble$, $4\pi r^3\rho^{\max}\coll(r)$
is the same as that shown in Figure~\ref{fig:Mcoll}(b)
(cf., eq.~\ref{eq:r3rhocolla}).
In the region with $t\coll(r)\gg t\Hubble$,
$4\pi r^3\rho^{\max}\coll(r)$ varies as $(a_*/\rsun)^2(\msun/m_*)$
[if $v\esc\ll\sigma(r)$] or $a_*/\rsun$ [if $v\esc\gg\sigma(r)$]
times $4\pi r^3\rho^{\max}\collsun(r)$ shown in
Figure~\ref{fig:Mcoll}(b) (cf., eq.~\ref{eq:r3rhocollb}).

\item For core galaxies, the mass involved in collisions
$M^{\max}\coll(t\Hubble)$,
which comes mainly from the region near the break radius $r\b$,
should be $a_*/\rsun$ times that shown in
Figure~\ref{fig:Mcoll}(a).

\item For power-law galaxies with $\gamma\ga 0.75$,
the mass involved in collisions
$M^{\max}\coll(t\Hubble)$ comes mainly from the region $r\simeq r\peak<r\H$ with
$t\coll(r\peak)\simeq t\Hubble$;
thus, according to equation (\ref{eq:tcollmstar}),
$r\peak$ should be $[(a_*/\rsun)^2(\msun/m_*)]^{1/(\gamma+1.5)}$
[if $\sigma(r)\gg v\esc$] or $(a_*/\rsun)^{1/(\gamma+0.5)}$
[if $\sigma(r)\ll v\esc$] times that shown in Figure~\ref{fig:Mcoll}(b),
and $M^{\max}\coll$
[$\propto r\peak^3\rho^{\max}\coll(r\peak)
\propto r\peak^3 n(r\peak)/t\Hubble\propto r\peak^{-\gamma+2}$]
is about $[(a_*/\rsun)^2(\msun/m_*)]^{(-\gamma+2)/(\gamma+1.5)}$
[if $\sigma(r)\gg v\esc$] or $(a_*/\rsun)^{(-\gamma+2)/(\gamma+0.5)}$
[if $\sigma(r)\ll v\esc$]
times $M^{\max}\collsun$ shown in Figure~\ref{fig:Mcoll}(a).
\end{itemize}

We now give two examples of the above changes.
We first assume that the galaxies are composed of low-mass stars with $m_*=0.1\msun$
and $a_*=0.16\rsun$[$\simeq(m_*/\msun)^\eta$, $\eta=0.8$, \citealt{KW90}].
The collision timescales $t\coll$ will increase by a factor of
$\sim4$ [$v\esc\ll\sigma(r)$] or $\sim6$ [$v\esc\gg\sigma(r)$]
compared to those in Figure~\ref{fig:tcoll}.
For core galaxies, 
the radii where the peaks of $4\pi r^3\rho^{\max}\coll(r)$ are located,
$r\peak$, are generally still around the break radii $r\b$,
but the stellar mass involved in collisions $M^{\max}\coll$ will decrease
by a factor of $\sim 0.2$ compared to Figure~\ref{fig:Mcoll}.
For power-law galaxies with $1.0\ga\gamma\ga0.75$,
$r\peak$ will decrease by a factor of $\sim 0.2$--0.6,
and so will the stellar masses involved in collisions $M^{\max}\coll$.
Similarly, we may also obtain the changes by assuming that the galaxies
are composed of high-mass stars with $m_*=10\msun$ and $a_*=6.3\rsun$.
The results are summarized in Table~1.

\begin{table*}
\begin{tabular}{lccccc}
\hline \hline
 &  \multicolumn{2}{c}{0.1$\msun$, 0.16$\rsun$} &
& \multicolumn{2}{c}{10$\msun$, 6.3$\rsun$} \\
\cline{2-3}\cline{5-6}
& core galaxy & power-law galaxy & & core galaxy & power-law galaxy \\
&            &  ($0.75\la\gamma\la1.0$) & &  & ($0.75\la\gamma\la1.0$) \\
\hline 
$r\peak/r\peak{_{,\odot}}$ & 1 & 0.2--0.6 & & 1 & 1.7--4.4\\
$M\coll^{\max}/M\collsun^{\max}$  & 0.2 & 0.2--0.6 & & 6 & 1.7--6.3\\
\hline
$t\coll/t\collsun$ & \multicolumn{2}{c}{4--6} & & \multicolumn{2}{c}{0.2--0.3}\\
\hline
\end{tabular}
\caption{Changes in the results shown in Figures~\ref{fig:tcoll} and
\ref{fig:Mcoll} (the stellar collision timescale $t\coll$, the total stellar
mass involved in collisions $M\coll^{\max}$, and the radius where most of the
stellar mass involved in collisions originates $r\peak$) by assuming that all
the stars have mass and radius lower than the solar values $(0.1\msun, 0.16\rsun)$
or higher than the solar values $(10\msun, 6.3\rsun)$.
See \S~\ref{sec:resBHgrowth}.}
\label{tab:tab1}
\end{table*}

\subsection{Generalization to a distribution of stellar masses and radii}
\label{subsec:gendis}

\noindent
A realistic stellar system is composed of stars with a distribution of
masses and radii, rather than identical stars.
To investigate this case, we assume that all stars are formed instantaneously
at the formation of the stellar system,
and use the following two forms of stellar initial mass function (IMF):
one is the Salpeter IMF \citep{S55},
\be
\Xi\Sp(m,t=0)=\cases{A\Sp(m/\msun)^{-2.35}
                                     & $0.08\le m/\msun\le 120$,  \cr
                    0                & otherwise; \cr
}
\label{eq:SpIMF}
\ee
and the other is the multi-power-law IMF (cf., \citealt{K02}),
\be
\Xi\multi(m,t=0)=A\multi\cases{
    (m/0.08\msun)^{-0.3} & $0.01\le m/\msun<0.08$,  \cr
    (m/0.08\msun)^{-1.3} & $0.08\le m/\msun<0.5$, \cr
    (0.5\msun/0.08\msun)^{-1.3}
    (m/0.5\msun)^{-2.3}  & $0.5\le m/\msun<120.0$, \cr
                    0                       & otherwise. \cr
}
\label{eq:multiIMF}
\ee
The constants $A\Sp$ and $A\multi$ in equations (\ref{eq:SpIMF}) and
(\ref{eq:multiIMF}) are determined by $\int_0^\infty\Xi(m,t=0)~\d m=1$.
The multi-power-law IMF gives a much flatter slope at the low-mass
end than the Salpeter IMF.
For stars with initial mass $m$, the lifetime on the main sequence is about
(eq.~9-3 in \citealt{BT87})
\be
t_{\rm MS}\simeq 10\Gyr(m/\msun)^{-2.5}.
\label{eq:tms}
\ee
In a stellar system with age $T\age\sim 10\Gyr$,
the mass of stars at the turn-off point $m\tf$ is about $1\msun$.
The present mass function of the remaining low-mass main-sequence stars
follows their initial mass function $\Xi(m,0)$ with $m\la m\tf$.
These stars are assumed to have radii $a\simeq(m/\msun)^{\eta}\rsun$ with
$\eta=0.8$ (p.208 in \citealt{KW90}).
Massive stars ($\ga 1\msun$) usually have lost a significant fraction of
their mass and become low-mass remnants (BHs, neutron stars or white dwarfs).
We assume that progenitors with mass $m>30\msun$ will become BHs with
mass $8\msun$, progenitors with mass $30\msun>m>8\msun$ will
become neutron stars with mass $1.5\msun$ and radius $10\km$,
and progenitors with mass $8\msun>m>1\msun$ will become white dwarfs
with mass $0.6\msun$ and radius $10^4\km$ (e.g., \citealt{MCD91}).
Besides stellar remnants, there also exist other, relatively short lived,
post-main-sequence
stars (such as red giants, horizontal branch stars or asymptotic giant branch
stars etc.) which have evolved from stars with initial
mass close to $m\tf$.

According to equation (\ref{eq:tcolli}), we have the collision timescale
of main-sequence stars with mass and radius $(m_i,a_i)$ as follows:
\be
t\colli(r,T\age)\propto \frac{1}
{n(r,T\age)\int\d m_j\d a_j~\xi(m_j,a_j,T\age) \Sigma_{ij}}. 
\label{eq:tcollmi}
\ee
Given the dynamically determined mass-to-light ratio 
(by normalizing to the central velocity dispersion based on spherical and
isotropic models fitted to the surface-brightness profile,
see \S~\ref{sec:galsmp})
and the observed surface brightness profile,
we have the following relation for the stellar number density obtained by
using the Eddington formula (eq. 4-140a in \citealt{BT87}) 
\be
n(r,T\age)\propto \frac{1}{\int_0^\infty\d m_j~m_j\Xi(m_j,T\age)},
\label{eq:nrTage}
\ee
which depends on the stellar mass function $\Xi(m_j,T\age)$.
Thus, using equations (\ref{eq:Sigmaij}), (\ref{eq:tcollmi}) and
(\ref{eq:nrTage}), we have
\begin{eqnarray}
& & \frac{t\colli(r,T\age)}{t\collsun(r,T\age)} \nonumber \\
& \simeq & \int_0^\infty\d m_j~(m_j/\msun)\Xi(m_j,T\age) \nonumber\\
& & \times \cases{
1/\int_{0}^{m\tf}\d m_j~\Xi(m_j,0)[(a_i+a_j)^2/(2\rsun)^2]
& if $\sigma(r)\gg v\escsun$ \cr
1/\int_{0}^{m\tf}\d m_j~\Xi(m_j,0)[(a_i+a_j)/2\rsun][(m_i+m_j)/2\msun]
& if $\sigma(r)\ll v\escsun$ \cr
},
\label{eq:tcollmisun}
\end{eqnarray}
where $t\collsun$ (see \S~\ref{sec:resBHgrowth} or Fig.~\ref{fig:tcoll})
is the collision timescale obtained by assuming that
all the stars have the solar mass and radius, and
$v\escsun=\sqrt{2G\msun/\rsun}$ is the escape velocity from stars with solar
mass and radius.
The two cases in equation (\ref{eq:tcollmisun}) are divided by
$v\escsun\sim \sigma(r)$ (cf., eq.~\ref{eq:Sigmaij}), because we have
$v_{ij}=\sqrt{2G(m_i+m_j)/(a_i+a_j)}\sim v\escsun$
for any collisions of two main-sequence stars
with masses $m_i$ and $m_j$ in the range 0.01--1$\msun$
[assuming $a\propto m^\eta$ with $\eta$(=0.8) being close to 1]
and we assume that the mean relative velocity of two colliding stars is
$\sqrt{\langle v\rel\rangle\langle v\rel^{-1}\rangle^{-1}}
\sim \langle v\rel^2\rangle^{1/2}=\sqrt{6}\sigma(r)\sim \sigma(r)$
(see eq.~\ref{eq:Sigmaij}).
In equation (\ref{eq:tcollmi}), 
we have ignored collisions with stellar remnants
(BHs, neutron stars or white dwarfs),
because the number of stellar remnants (evolved from massive stars) is small
compared to that of main-sequence stars
(note that mass segregation is ignored)
and also the radius of stellar remnants is not large ($<1\rsun$).
We have also ignored collisions with other post-main-sequence stars
(e.g., red giants) since the number of these stars is also small.
Note that the large radius of giants (e.g., $\sim10^2\rsun$) will increase
the collision rates of main-sequence stars,
but collisions with red-giant envelopes generally result in little mass
loss from either star
because the amount of mass contained in the greatly extended envelopes of
giants is typically a small fraction of a solar mass and the overlapping area
of collisions is quite small relative to the total collision cross section.
In addition, as argued by \citet{MCD91}, giants undergo significant mass
loss in a short amount of time anyway because of stellar winds,
and it makes not much difference whether the mass comes from winds
or collisions in the sense of providing raw material for the BH growth.

\begin{figure}
\begin{center}
\includegraphics[width=0.8\textwidth,angle=0]{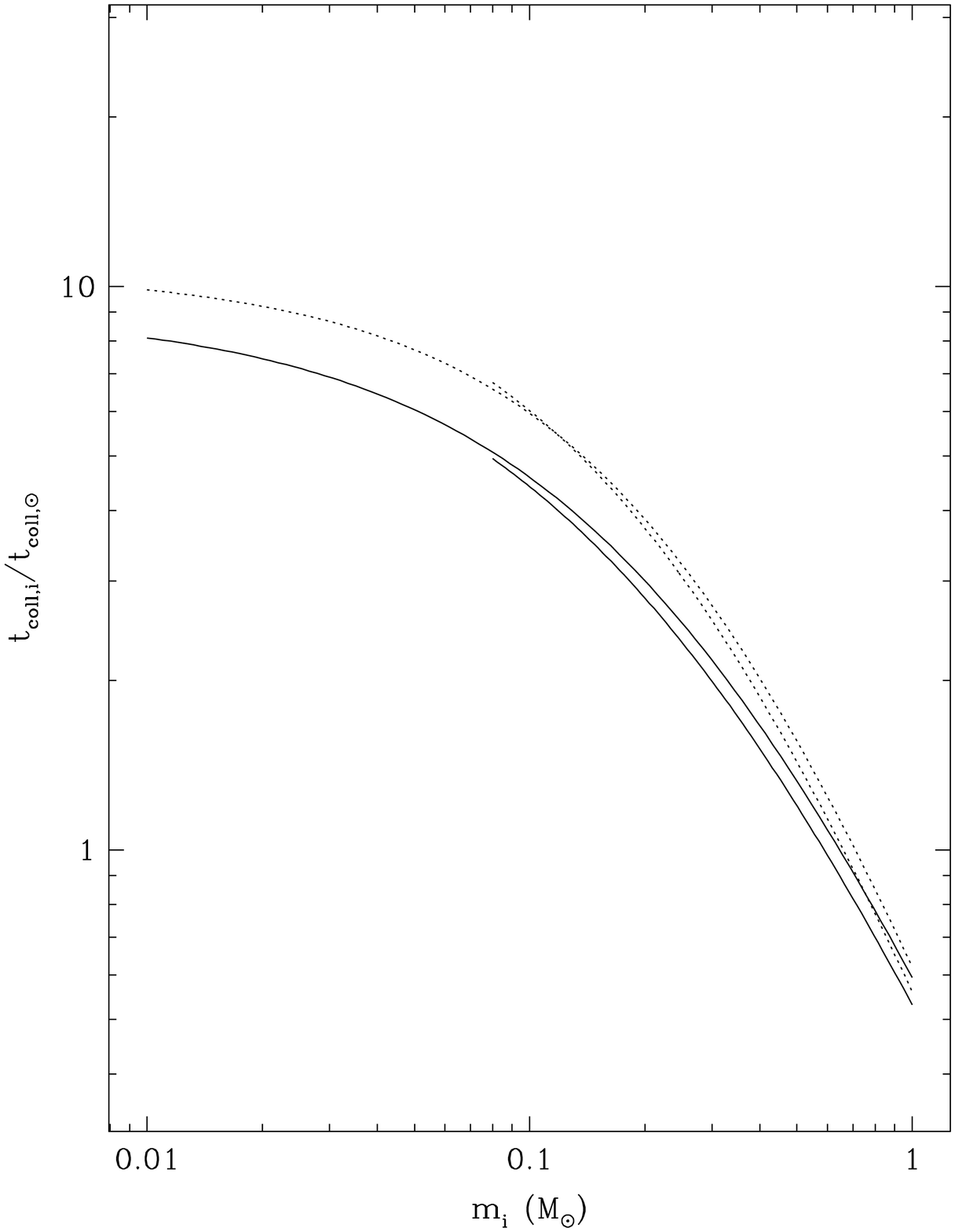}
\caption{The ratio of collision timescales $t\colli/t\collsun$
as a function of stellar mass $m_i$.
The quantity $t\colli$ (see eq.~\ref{eq:tcolli} or \ref{eq:tcollmisun})
is the collision timescale of stars with mass $m_i$ obtained by assuming that
the stellar system has a distribution of stellar masses and radii
and ignoring collisions with post-main-sequence stars and stellar remnants;
and $t\collsun$ (see Fig.~\ref{fig:tcoll})
is the collision timescale of stars
by assuming that stars in the stellar system have solar mass and radius.
Note that $t\colli$ at $m_i=\msun$ is not necessarily equal to $t\collsun$.
The dotted curves represent the results for $\sigma(r)\ll v\escsun$,
and the solid curves represent the results for $\sigma(r)\gg v\escsun$.
The curves ending at $0.08\msun$ are obtained by assuming
the Salpeter IMF (eq.~\ref{eq:SpIMF}),
and those ending at $0.01\msun$ are obtained by assuming
the multi-power-law IMF (eq.~\ref{eq:multiIMF}).
The turn-off stars are assumed to have a mass $m\tf=1\msun$.
}
\label{fig:tcollmi}
\end{center}
\end{figure}

Using equation (\ref{eq:tcollmi}) and assuming that the stellar radius
$a_i=(m_i/\msun)^{-0.8}\rsun$ and the mass of stars at the turn-off point
$m\tf=1\msun$,
we obtain the collision timescales $t\colli$ as a function of stellar mass
$m_i$.
The ratio of collision timescales $t\colli/t\collsun$ (eq.~\ref{eq:tcollmisun})
as a function of stellar mass $m_i$ is shown in Figure~\ref{fig:tcollmi}
(see $t\collsun$ in Fig.~\ref{fig:tcoll}).
We show the results obtained both by assuming the Salpeter IMF
(eq.~\ref{eq:SpIMF}) and by assuming the multi-power-law IMF
(eq.~\ref{eq:multiIMF}).
As seen from Figure~\ref{fig:tcollmi}, the ratio 
$t\colli/t\collsun$ increases from $\sim 0.6$ to $\sim 7$ (or to $\sim 9$)
with decreasing stellar mass $m_i$, from $m\tf$ to $0.08\msun$ for the
Salpeter IMF (or to $0.01\msun$ for the multi-power-law IMF),
and the difference of the ratios between the two IMFs for
$\sigma(r)\ll v\escsun$ and for $\sigma(r)\gg v\escsun$ is less than 30 percent.
For the whole population of main-sequence stars, we use equation
(\ref{eq:tcoll}) and the collision timescale $t\colli$
shown in Figure~\ref{fig:tcollmi}
to obtain the total collision timescale $t\coll$, which
is about 3--4 times the timescales $t\collsun$ shown in Figure~\ref{fig:tcoll}.

\begin{figure}
\begin{center}
\includegraphics[width=0.8\textwidth,angle=0]{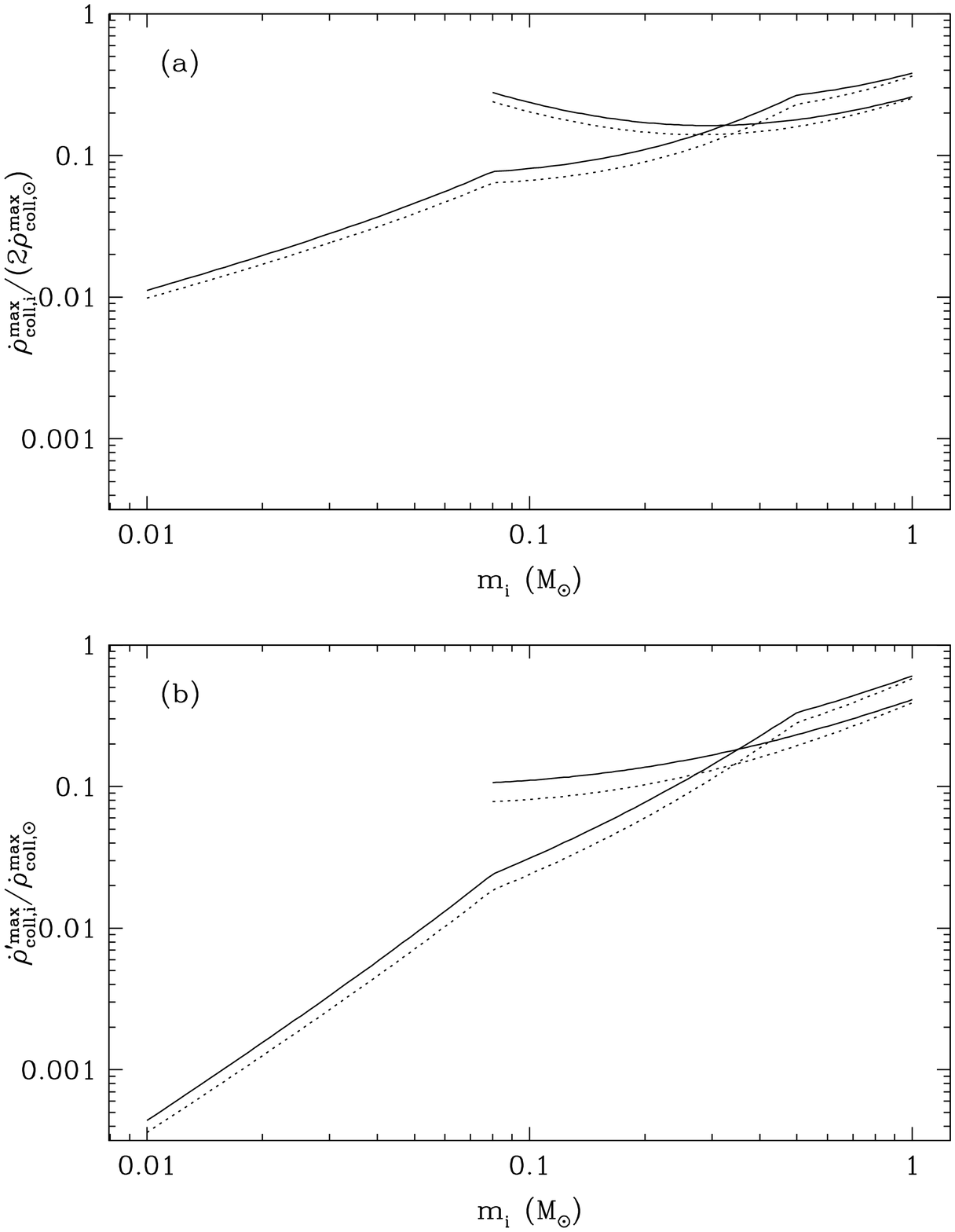}
\caption{Panel(a):
The ratio of $\dot\rho\colli^{\max}/(2\dot\rho\collsun^{\max})$
as a function of stellar mass $m_i$
(cf., eqs.~\ref{eq:dotrhocolli} and \ref{eq:dotrhocollisun}).
Panel (b):
The ratio of $\dot\rho\colli\pmax/\dot\rho\collsun^{\max}$
as a function of stellar mass $m_i$
(cf., eqs.~\ref{eq:dotrhocollij} and \ref{eq:dotrhocollijsun}).
The quantity $\dot\rho\colli^{\max}\d \ln m_i$
is the mass density of main-sequence stars involved in collisions with
main-sequence stars with mass in the range
$\ln m_i\rightarrow \ln m_i+\d \ln m_i$ per unit time;
the quantity $\dot\rho\colli\pmax\d \ln m_i$ is the mass density
of main-sequence stars with mass in the range
$\ln m_i\rightarrow \ln m_i+\d \ln m_i$
involved in collisions per unit time;
and the quantity $\dot\rho\collsun^{\max}$ is the mass density of main-sequence
stars involved in collisions per unit time that is obtained
by assuming that all the stars
in the stellar system have solar mass and radius.
The ratios $\dot\rho\colli^{\max}/(2\dot\rho\collsun^{\max})$ and
$\dot\rho\colli\pmax/\dot\rho\collsun^{\max}$ are not necessarily 1
at $m_i=\msun$.
The curves ending at $0.08\msun$ are obtained by assuming the
Salpeter IMF (eq.~\ref{eq:SpIMF}) and those ending at
$0.01\msun$ are obtained by assuming the multi-power-law IMF
(eq.~\ref{eq:multiIMF}).
The dotted curves represent the results for $\sigma(r)\ll v\escsun$,
and the solid curves represent the results for $\sigma(r)\gg v\escsun$.
}
\label{fig:mcollmi}
\end{center}
\end{figure}

According to equation (\ref{eq:dotrhocoll}),
we have the total mass density of main-sequence stars involved in collisions
with main-sequence stars with the natural logarithm of their masses
in the range $\ln m_i\rightarrow \ln m_i+\d\ln m_i$
per unit time given by:
\begin{eqnarray}
\dot\rho\colli^{\max}(r,T\age)\d\ln m_i
 & \propto & \d m_i\int_0^{m\tf}\d m_j~(m_i+m_j){\cal R}_{ij}(r,T\age) \nonumber
 \\
 & \propto & \d m_i\int_0^{m\tf}\d m_j~(m_i+m_j)\Xi(m_i,0)\Xi(m_j,0)
n^2(r,T\age)\Sigma_{ij};
\label{eq:dotrhocolli}
\end{eqnarray}
and using the similar derivation to obtain equation (\ref{eq:tcollmisun}),
we have the ratio
\begin{eqnarray}
\frac{\dot\rho\colli^{\max}(r,T\age)}{2\dot\rho\collsun^{\max}(r,T\age)}
\simeq m_i\Xi(m_i,0)\left[\int_0^\infty\d m_j~(m_j/\msun)\Xi(m_j,T\age)\right]^{-2}
\qquad\qquad\qquad\qquad\qquad\qquad & & \nonumber \\
\times \cases{
\int_{0}^{m\tf}\d m_j~\Xi(m_j,0)[(a_i+a_j)/(2\rsun)]^2[(m_i+m_j)/(2\msun)] & $ \sigma(r)\gg v\escsun$ \cr
\int_{0}^{m\tf}\d m_j~\Xi(m_j,0)[(a_i+a_j)/(2\rsun)][(m_i+m_j)/(2\msun)]^2 & $\sigma(r)\ll v\escsun$ \cr
}, & &
\label{eq:dotrhocollisun}
\end{eqnarray}
where $\dot\rho\collsun^{\max}$ (cf., eq.~\ref{eq:dotrhocoll}) is the total
mass density of main-sequence stars involved in collisions per unit time
obtained by assuming that all the stars in the stellar system have solar
mass and radius.
We put the factor ``$1/2$'' in front of 
$\dot\rho\colli^{\max}(r,T\age)/\dot\rho\collsun^{\max}(r,T\age)$
because $\int_0^{m\tf}\dot\rho\colli^{\max}(r,T\age)~\d\ln m_i$ gives
twice of the mass involved in collisions
(see also the factor ``$1/2$'' in eq.~\ref{eq:dotrhocoll}).
We also have the total mass density
of main-sequence stars with the natural logarithm of their masses
in the range $\ln m_i\rightarrow \ln m_i+\d \ln m_i$
involved in collisions per unit time:
\begin{eqnarray}
\dot\rho\colli\pmax(r,T\age)\d \ln m_i
 &\propto & \d m_i\int_0^{m\tf}\d m_j~m_i{\cal R}_{ij}(r,T\age) \nonumber \\
 &\propto & \d m_i\int_0^{m\tf}\d m_j~m_i\Xi(m_i,0)\Xi(m_j,0)
n^2(r,T\age)\Sigma_{ij},
\label{eq:dotrhocollij}
\end{eqnarray}
and
\begin{eqnarray}
\frac{\dot\rho\colli\pmax(r,T\age)}{\dot\rho\collsun^{\max}(r,T\age)}
\simeq (m_i/\msun)m_i\Xi(m_i,0)
\left[\int_0^\infty\d m_j~(m_j/\msun)\Xi(m_j,T\age)\right]^{-2}
\qquad\qquad\qquad\qquad & & \nonumber \\
\times \cases{
\int_{0}^{m\tf}\d m_j~\Xi(m_j,0)[(a_i+a_j)/(2\rsun)]^2 & $ \sigma(r)\gg v\escsun$ \cr
\int_{0}^{m\tf}\d m_j~\Xi(m_j,0)[(a_i+a_j)/(2\rsun)][(m_i+m_j)/(2\msun)] & $\sigma(r)\ll v\escsun$ \cr
}. & &
\label{eq:dotrhocollijsun} 
\end{eqnarray}
As in equations (\ref{eq:tcollmi}) and (\ref{eq:tcollmisun}),
collisions with stellar remnants and other post-main-sequence stars are ignored
in equations (\ref{eq:dotrhocolli})--(\ref{eq:dotrhocollijsun}).
We use equations (\ref{eq:dotrhocolli})--(\ref{eq:dotrhocollijsun}) to
obtain $\dot\rho\colli^{\max}(r,T\age)$ and $\dot\rho\colli\pmax(r,T\age)$
as a function of stellar mass $m_i$.
The ratios of
$\dot\rho\colli^{\max}/(2\dot\rho\collsun^{\max})$ and
$\dot\rho\colli\pmax/\dot\rho\collsun^{\max}$
as a function of stellar mass $m_i$
for both the cases $\sigma(r)\ll v\escsun$ (dotted lines)
and $\sigma(r)\gg v\escsun$ (solid lines)
are shown in Figure~\ref{fig:mcollmi}(a) and (b), respectively.
We also show the results obtained both by assuming the Salpeter IMF
(eq.~\ref{eq:SpIMF}) and by assuming the multi-power-law IMF
(eq.~\ref{eq:multiIMF}).
In Figure~\ref{fig:mcollmi}(a),
the curves of $\dot\rho\colli^{\max}/(2\dot\rho\collsun^{\max})$
obtained from the Salpeter IMF (with the low-mass end at $0.08\msun$)
show a minimum at an intermediate mass between 0.08--$1\msun$
and increase from the minima to both the the high-mass end ($\sim 1\msun$)
and the low-mass end ($\sim 0.08\msun$),
which reflects the fact that the stellar mass involved in collisions
comes mainly from collisions between high-mass stars and low-mass stars;
and the curves of $\dot\rho\colli^{\max}/(2\dot\rho\collsun^{\max})$
obtained from the multi-power-law IMF (with the low-mass
end at $0.01\msun$) increase with increasing stellar mass $m_i$, 
which reflects the fact that the stellar mass involved in collisions
comes mainly from collisions with high-mass stars
since the multi-power-law IMF gives a flatter slope at the low-mass end than
the Salpeter IMF.
Figure~\ref{fig:mcollmi}(b) shows that
$\dot\rho\colli\pmax/\dot\rho\collsun^{\max}$ obtained from
both the Salpeter IMF and the multi-power-law IMF increase with increasing
stellar mass $m_i$, which reflects that the stellar mass involved in
collisions comes mainly from high-mass stars, especially for the
multi-power-law IMF since
the slope of corresponding curves in Figure~\ref{fig:mcollmi}(b) is steeper.
For core galaxies, the total mass of main-sequence stars involved in
collisions comes mainly from galactic radii $r\sim r\b$
[where $t\colli(r)\gg t\Hubble$ and $\sigma(r)\ll v\escsun$],
which is found to be
$\sim\int_0^{m\tf}\dot\rho\colli^{\max}~\d \ln m_i/(2\dot\rho\collsun^{\max})
\simeq 0.4$--0.5
times the result obtained by assuming that the stellar systems have
identical stars with solar mass and radius.
For power-law galaxies, using equation (\ref{eq:Mcolldis}),
we find that the total stellar mass involved in collisions
$M^{\max}\coll$ is smaller than $M^{\max}\collsun$
(see \S~\ref{sec:resBHgrowth} or Fig.~\ref{fig:Mcoll})
by a factor of 0.4--0.7.

As shown above, after the generalization to a distribution of stellar masses
and radii, the total stellar masses involved in collisions decrease compared
to those shown in Figure~\ref{fig:Mcoll}(a).
In addition, note that not all the collisions result in disruption of stars
and not all the gas released by collisions can be accreted onto BHs.
Hence, the conclusion obtained from Figure~\ref{fig:Mcoll}
that the mass released by stellar collisions is not enough to
account for the central
BH growth (especially in bright core galaxies) will not be changed,
by more realistic models.

Mass segregation is ignored in the analysis above.
This assumption affects little on our result of stellar mass involved in
collisions at least for all the core galaxies and most of the power-law
galaxies for the following reasons.
Our calculation shows that the relaxation timescales of stellar remnants
of massive stars (BHs and neutron stars) are higher than the Hubble time
for all the core galaxies and are higher than the
Hubble time at $r\ga r\peak{_{,\odot}}$ for most of the power-law galaxies
[only for about four power-law galaxies, the relaxation timescales of
the stellar BHs ($8\msun$) are smaller than or approximately equal to
the Hubble time at the region $r\simeq r\peak{_{,\odot}}$.].

\subsection{Material sources for growth of central BHs} \label{subsec:grow}

\noindent
As mentioned in \S~\ref{sec:intro}, gaseous material for the growth of
central BHs
in an isolated stellar system may come from stars through three mechanisms
(e.g., \citealt{MCD91}): tidal disruption (or being swallowed whole by
central BHs) of stars (mostly on elongated radial orbits), stellar collisions,
and stellar evolution.
In this subsection, we will compare these three contributors to BH growth.

\begin{figure}
\begin{center}
\includegraphics[width=0.8\textwidth,angle=0]{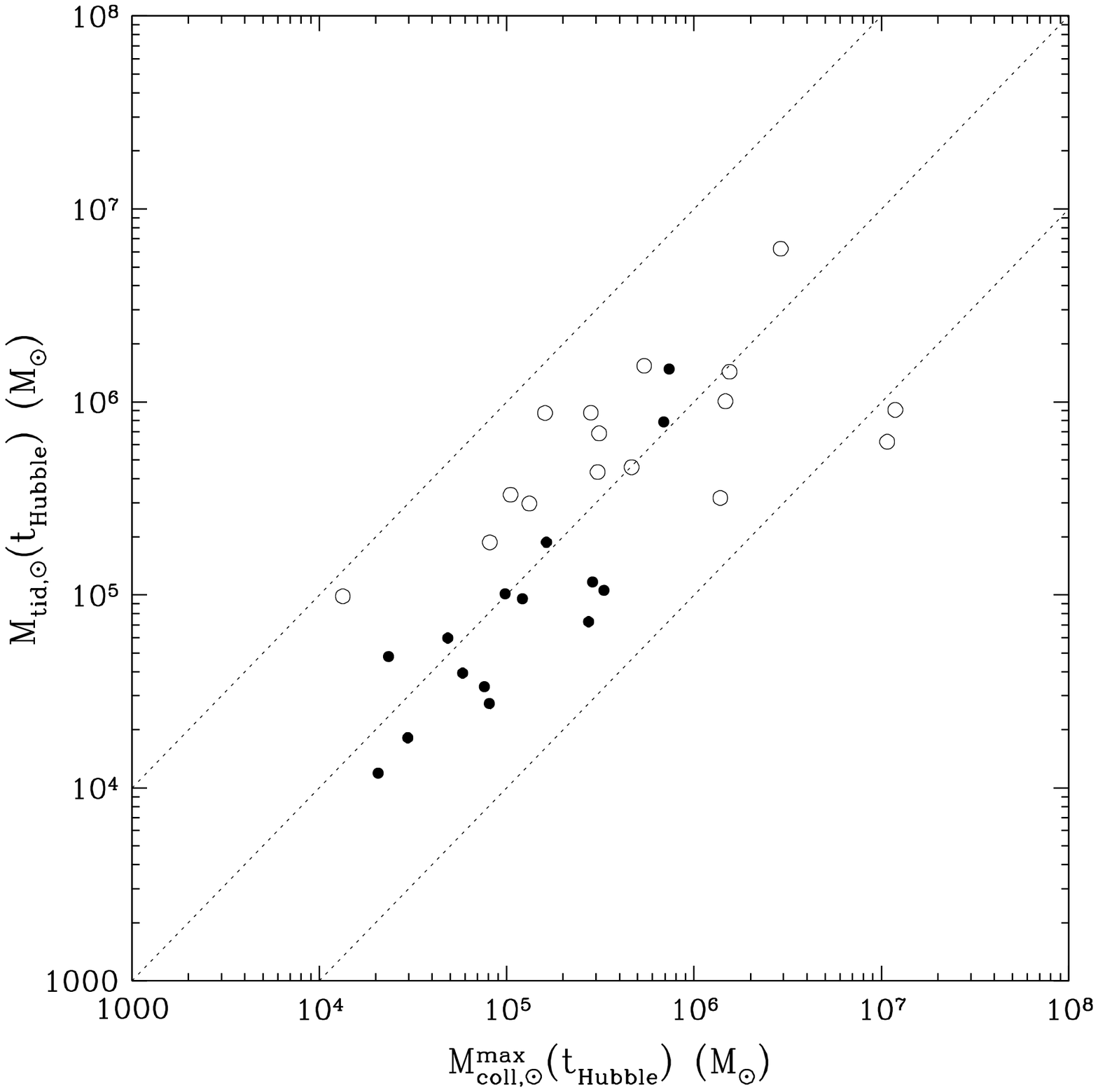}
\caption{Stellar mass tidally disrupted or swallowed whole by central BHs
$M\tidsun(t\Hubble)$ versus stellar mass involved in collisions
$M\collsun^{\max}(t\Hubble)$ over a Hubble time (see Fig.~\ref{fig:Mcoll}a).
All stars are assumed to have the solar mass and radius.
Dotted lines are the reference lines for $M\tid/M^{\max}\coll=0.1,1,10$.
Solid circles represent core galaxies and open circles represent
power-law galaxies.
$M\tidsun$ and $M\collsun^{\max}$ are generally comparable.
See discussion in \S~\ref{subsec:grow}.
}
\label{fig:mcolltid}
\end{center}
\end{figure}

In a stellar system with a central BH, stars that come within a distance
$r\tid\sim (M_\bullet/m_*)^{1/3}a_*$ of the central BH will be tidally
disrupted by the BH if $r\tid\ga r\Sch$ (where $r\Sch\equiv 2GM_\bullet/c^2$
is the Schwarzschild radius) or swallowed whole by the BH if $r\tid\la r\Sch$.
Here, we do not distinguish the above two cases and
simply call the total mass of stars which may come within a distance of
$r=\max(r\tid, r\Sch)$ as the tidally disrupted mass $M\tid$.
We use the same method as Magorrian \& Tremaine (1999) to obtain
$M\tid$ in realistic galaxies,
and the galaxies are assumed to be spherical since flattening or triaxiality
of galaxies does not significantly change the results \citep{MT99}.
Figure~\ref{fig:mcolltid} shows the tidally disrupted stellar mass
($M\tidsun$) versus the stellar mass involved in collisions ($M\collsun^{\max}$)
over a Hubble time;
in this Figure, the stellar systems are assumed to be composed of identical
stars with solar mass and radius. 
As seen from Figures~\ref{fig:Mcoll} and \ref{fig:mcolltid},
$M\tidsun$ and $M\collsun^{\max}$ are generally comparable. 
After the generalization to a distribution of stellar masses and radii
(i.e., the Salpeter IMF in eq.~\ref{eq:SpIMF} or
the multi-power-law IMF in eq.~\ref{eq:multiIMF}),
$M\tid$ will increase by a factor of $\sim$2--3 (cf., \citealt{MT99}),
while $M\coll^{\max}$ will be reduced by a factor of $\sim 0.4$--0.7.
Considering that not all the collisions result in stellar disruption
and not all the gas released by collisions can be accreted onto BHs,
tidal disruption is likely to contribute more mass to the central BH growth
than stellar collisions.
\citet{MT99} show that tidal disruption contributes significantly 
to the present BH mass only in faint galaxies $\la 10^9\Lsun$.
For bright galaxies, our study in this paper shows that neither tidal
disruption nor stellar collisions can contribute significantly to their present
central BH masses.

Using the simple model of stellar evolution described at the beginning
of \S~\ref{subsec:gendis},
we may obtain the mass loss caused by stellar evolution if all stars are
formed instantaneously at the formation of the stellar system.
The mass loss may comes either from stellar winds (mainly during giant
stages) or from the mass lost by Supernovae II.
The total mass loss amounts to about 21 percent of the initial total
stellar mass in the initial $1\Gyr$ and about 7 percent during the interval
from $1\Gyr$ to $10\Gyr$ for the Salpeter IMF.
The corresponding numbers for the multi-power-law IMF are 32\% and 10\%.
(For both of the IMFs, more than forty percent of the mass loss in the initial
$1\Gyr$ comes from stars with initial mass in the range 2.5--$8\msun$ or
mass lost by stellar winds.)
Therefore, mass loss during stellar evolution at the early stage of
galaxy evolution is a possible source to feed central BHs if
part of the lost mass can lose angular momentum and sink to galactic nuclei.

A final possibility for the material source is that
galaxy encounters may concentrate gas in the centers of
interacting galaxies to feed BHs (see simulations in \citealt{BH96}).
The gas infall may fuel pre-existing central BHs;
or galaxy encounters may trigger a central star burst,
which could lead to nuclear activity.
For both of the above possibilities,
massive BHs form or grow rapidly, early in the universe's history
(e.g., at redshift $z\sim$2--3 when galaxy encounters mostly occur
in the hierarchical galaxy formation model).
This scenario is consistent with one of the arguments made by \citet{YT02}
that growth of high-mass BHs ($>10^8\msun$) comes
mainly from accretion during optically bright QSO phases
(which is obtained by studying the relation between the local BH mass function
and the QSO luminosity function).

\section{Effects of stellar collisions on colors of galactic centers}
\label{sec:galcolor}

\noindent
As mentioned in \S~\ref{sec:intro}, stellar collisions
affect the luminosity properties and colors in galactic centers.
Stellar collisions may disrupt one or both of the colliding stars to
decrease the total luminosity,
or the two colliding stars may merge to become a new star with
luminosity properties or colors different from its parent stars.
In this section, we will first give an general analysis of the effects
of stellar collisions on color gradients, and then
study these effects in nearby galactic centers,
especially two galaxies in the Local Group: M32 and M31.

\subsection{Color gradients caused by stellar collisions}
\label{subsec:colorgrad}

\noindent
Color gradients in a galaxy are usually caused by the variation of its
stellar population with galactic radius.

If there are no collisions in the stellar system,
the surface brightness at color band $C$ and at 
projected radius $R$, is given by: 
\begin{eqnarray}
I_{0,C}(R,T\age)
&=&\int\d z\int\d m\d a~\xi_0(m,a,T\age)n(r,T\age) L_C(m,a,T\age) \label{eq:I0}\\
&=& J_{0,C}(R,T\age)Rf_0(R).
\label{eq:I0a}
\end{eqnarray}
Here $r=\sqrt{R^2+z^2}$, $\xi_0(m,a,T\age)$ is the expected stellar mass
and radius function at the present time obtained by assuming an IMF and then
only considering the effect of stellar evolution,
$L_C(m,a,T\age)$ is the current luminosity emitted by a star with
mass and radius $(m,a)$ at color band $C$
(in units of the solar $C$-band luminosity $L_{\odot,C}$),
$J_{0,C}(r,T\age)$ is the luminosity density at galactic radius $r$
defined by
\be
J_{0,C}(r,T\age)\equiv\int\d m\d a~\xi_0(m,a,T\age)n(r,T\age) L_C(m,a,T\age)
\label{eq:J0}
\ee
and $f_0$ is a factor reflecting the projection effect of the
surface brightness defined by:
\be
f_0(R)\equiv\frac{\int_{-\infty}^\infty n(r,T\age)\d z}{n(R,T\age)R}.
\label{eq:f0}
\ee
The color index of the surface brightness (at color bands $C_1$ and $C_2$)
at projected radius $R$ is given by:
\be
\mu_{0,C_1}(R)-\mu_{0,C_2}(R)=-2.5\log[I_{0,C_1}(R)/I_{0,C_2}(R)]
+(M_{\odot,C_1}-M_{\odot,C_2}),
\label{eq:colorgrd0}
\ee
where $M_{\odot,C_i}$ is the solar absolute magnitude
at color band $C_i$.
We will call the surface brightness and the color in equations (\ref{eq:I0})
and (\ref{eq:colorgrd0}) the ``original'' surface brightness and
color, that is, the brightness and color that would be present in the
absence of stellar collisions.

If two stars with masses and radii $(m_i,a_i)$ and $(m_j,a_j)$ collide
at time $t'$, we define $L_{ij,C}(t,t')$ ($t\ge t'$) as the $C$-band
luminosity of the collision product at time $t$.
The luminosity and color of collision products depend on the types of
colliding stars, their masses and radii, their relative velocity and
impact parameter etc.
If both of the two stars are disrupted by the collision, we have
$L_{ij,C}(t,t')=0$;
if the two stars merge to be a new star, $L_{ij,C}(t,t')$ is just the
luminosity of the new merged star;
and if one or both of the two stars survive the collision, $L_{ij,C}(t,t')$
represents the total luminosity of the surviving star(s) after the collision.
Using the collision rate ${\cal R}_{ij}(r,t')$ in equation (\ref{eq:calRij}),
the change of the surface brightness at color band $C$ caused by collisions
of two stars with total mass larger than $m\ij$ is given by:
\begin{eqnarray}
& & I\collmij{_{,C}}(R,T\age) \nonumber \\
& = & {1\over2}\int^{m\tf}_{\max(m\ij-m\tf,0)}\d m_i\int^{m\tf}_{\max(m\ij-m_i,0)}\d m_j \int\d a_i\int \d a_j\int\d z
\int_0^{T\age}\d t'~{\cal R}_{ij}(r,t')
\nonumber \\
& & \qquad \qquad \qquad \times [\langle L_{ij,C}(T\age,t',r)\rangle-L_C(m_i,a_i,T\age)-L_C(m_j,a_j,T\age)],
\label{eq:Icollij}
\end{eqnarray}
where $\langle L_{ij,C}(T\age,t',r)\rangle$ is the average present luminosity
of the collision product of two stars with masses and radii $(m_i,a_i)$
and $(m_j,a_j)$ colliding at time $t'$ and at galactic radius $r$.
We have the factor ``1/2'' in front of the integration in equation
(\ref{eq:Icollij}) for the same reason given for
equation (\ref{eq:dotrhocoll}), and we have
\be
{1\over2}\int^{m\tf}_{\max(m\ij-m\tf,0)}\d m_i\int^{m\tf}_{\max(m\ij-m_i,0)}\d m_j
=\int^{m\tf}_{\max(m\ij-m\tf,0)}\d m_i\int^{m\tf}_{\max(m\ij-m_i,m_i)}\d m_j.
\ee
The dependence of the average luminosity
$\langle L_{ij,C}(T\age,t',r)\rangle$ on $r$ is due
to the dependence of the collision outcome on the relative velocity of two
colliding stars and the dependence of the relative velocity on $r$.
In equation (\ref{eq:Icollij}), we have ignored the radial motion of the
collision products, because the radial motion of the collision
products is expected to dilute the color gradients caused by collisions; 
thus, the color gradient obtained from equation (\ref{eq:Icollij})
is an overestimate.
We will mainly analyze the color gradient in the region
with $t\coll(r)\gg T\age$, where the stellar distribution and population
[i.e., the stellar density $n(r)$ and the stellar mass and radius function
$\xi(m,a)$] evolve rather slowly,
and we ignore collisions with collision products.
Once the initial rapid stellar evolution is complete
[massive stars have evolved into stellar remnants and the main-sequence stars
have low masses (e.g., $\la 1\msun$) with long lifetime
in the main-sequence phase],
in the slowly evolved region [where $t\coll(r)\gg T\age$], we have
${\cal R}_{ij}(r,t)\simeq {\cal R}_{ij}(r,0)\simeq {\cal R}_{ij}(r,T\age)$
in equation (\ref{eq:Icollij}), i.e.,
\begin{eqnarray}
& & I\collmij{_{,C}}(R,T\age) \nonumber \\
& = & {1\over2}\int^{m\tf}_{\max(m\ij-m\tf,0)}\d m_i\int^{m\tf}_{\max(m\ij-m_i,0)}\d m_j\int\d a_i\int\d a_j\int\d z~{\cal R}_{ij}(r,T\age) \nonumber \\
& &\times
\left[\int_0^{T\age} \langle L_{ij,C}(T\age,t',r)\rangle \d t'
-L_C(m_i,a_i,T\age)T\age-L_C(m_j,a_j,T\age)T\age\right].
\label{eq:Icollija}
\end{eqnarray}
Thus, the color index at projected radius $R$ is given by:
\begin{eqnarray}
\mu_{C_1}(R)-\mu_{C_2}(R)=-2.5\log\left[\frac{I_{0,C_1}(R)+I\coll{_{,>0,C_1}}(R)}{I_{0,C_2}(R)+I\coll{_{,>0,C_2}}(R)}\right]
+(M_{\odot,C_1}-M_{\odot,C_2}),
\label{eq:colorgrd0coll}
\end{eqnarray}
and the change of the color index caused by stellar collisions is given by:
\begin{eqnarray}
& & \Delta\mu\coll{_{,C_1-C_2}}(R)=[\mu_{C_1}(R)-\mu_{C_2}(R)]-[\mu_{0,C_1}(R)-\mu_{0,C_2}(R)] \label{eq:Deltamucoll}.
\end{eqnarray}
The contribution to the change of the color index
from collisions between two stars with total mass $m\ij$
can be seen from the following analysis.
According to equations (\ref{eq:colorgrd0coll}) and (\ref{eq:Deltamucoll}),
we have
\begin{eqnarray}
& & \Delta\mu\coll{_{,C_1-C_2}}(R) \nonumber \\
& = & -2.5\log\left[1+\int\d \ln m\ij~\frac{-\d I\collmij{_{,C_1}}(R)/\d\ln m\ij}{I_{0,C_1}(R)}\right] \nonumber \\
& &\hphantom{xxxxxxxxxxxx}+2.5\log\left[1+\int\d \ln m\ij~\frac{-\d I\collmij{_{,C_2}}(R)/\d \ln m\ij}{I_{0,C_2}(R)}\right]
\label{eq:colorgrdstr}\\
& \simeq & -\frac{2.5}{\ln 10}\int\d \ln m\ij
\left[\frac{-\d I\collmij{_{,C_1}}(R)/\d \ln m\ij}{I_{0,C_1}(R)}-\frac{-\d I\collmij{_{,C_2}}(R)/\d \ln m\ij}{I_{0,C_2}(R)}\right] \label{eq:colorgrd}\\
& \simeq & \int\d \ln m\ij~\frac{-\d I\collmij{_{,C_1}}(R)/\d \ln m\ij}{I_{0,C_1}(R)}
\Delta\mu\coll{_{,m\ij,C_1-C_2}}(R), \label{eq:colorgrdcollappro}
\end{eqnarray}
where
\begin{eqnarray}
& & \frac{\d I\collmij{_{,C_1}}(R)}{\d\ln m\ij} \nonumber \\
& = & -{m\ij \over2} \left.\int^{m\tf}_{\max(m\ij-m\tf,0)}\d m_i
\int\d a_i\int\d a_j\int\d z~{\cal R}_{ij}(r,T\age)\right|_{m_j=m\ij-m_i}  \nonumber \\
& & \times
\left[\int_0^{T\age} \langle L_{ij,C}(T\age,t',r)\rangle \d t'
-L_C(m_i,a_i,T\age)T\age-L_C(m_j,a_j,T\age)T\age\right],
\label{eq:dIcollijaBS}
\end{eqnarray}
\be
\Delta\mu\coll{_{,m\ij,C_1-C_2}}(R)=
[\mu\coll{_{,m\ij,C_1}}(R)-\mu\coll{_{,m\ij,C_2}}(R)]-[\mu_{0,C_1}(R)-\mu_{0,C_2}(R)],
\ee
\be
\mu\coll{_{,m\ij,C_1}}(R)-\mu\coll{_{,m\ij,C_2}}(R)=
-2.5\log\left[\frac{\d I\collmij{_{,C_1}}(R)/\d \ln m\ij}{\d I\collmij{_{,C_2}}(R)/\d \ln m\ij}\right]+(M_{\odot,C_1}-M_{\odot,C_2}).
\label{eq:colorgrdcoll}
\ee
In equations (\ref{eq:colorgrd}) and (\ref{eq:colorgrdcollappro}),
we use the relation $\ln(x)=x-1$ if $x\rightarrow 1$ where
$\ln x =\Delta\mu\coll{_{,m\ij,C_1-C_2}}/2.5$.
Equations (\ref{eq:colorgrd}) and (\ref{eq:colorgrdcollappro}) are valid for
$I\coll{_{,>0,C_1}}(R)/I_{0,C_1}(R)\ll 1$ and
$I\coll{_{,>0,C_2}}(R)/I_{0,C_2}(R)\ll 1$;
and for equation (\ref{eq:colorgrdcollappro}), we also need
$|\Delta\mu\coll{_{,m\ij,C_1-C_2}}(R)|\ll 2.5$.
As seen from equation (\ref{eq:colorgrdcollappro}),
for the collisions of two stars with the natural logarithm of their total
mass in the range $\ln m\ij\rightarrow \ln m\ij+\d \ln m\ij$,
whether the change of colors caused by collisions is significant is determined
by two factors:
(i) the ratio of the change of surface brightness caused by
collisions to the original surface brightness,
$\d \ln m\ij(\d I\collmij{_{,C_1}}/\d \ln m\ij)/I_{0,C_1}$,
(ii) the difference between the color index of the change of surface
brightness caused by collisions
$\mu\coll{_{,m\ij,C_1}}-\mu\coll{_{,m\ij,C_2}}$
and the original color index $\mu_{0,C_1}-\mu_{0,C_2}$ of the stellar
system (i.e., $\Delta\mu\coll{_{,m\ij,C_1-C_2}}$).

In this paper, we focus on studying the effects on the color gradient
in visible bands caused by collisions between main-sequence stars.
We ignore the effects of collisions with post-main-sequence stars
because the outcome (especially luminosity and color properties) of these
collisions is not well understood and detailed studies of collisions with
post-main-sequence stars
(e.g., \citealt{BD99}) are beyond the scope of this work.
We also ignore collisions of compact stellar remnants with main-sequence
stars, post-main-sequence stars or even with other compact remnant stars,
which usually form ``exotic'' objects and occur at low rates.
We do not intend to study the color gradients in $UV$, far-$UV$ or other
invisible bands in realistic galaxies, because
the luminosities in these bands are more sensitive to
post-main-sequence stars (e.g., giants or horizontal branch stars) etc.

\subsubsection{Luminosity of collision products of main-sequence stars}
\label{sec:BS}

\noindent
The outcome of collisions between main-sequence stars has been investigated
by smoothed particle hydrodynamics simulations
[e.g., see \citet{Sills97,Sills01} for collisions in globular
clusters, and \citet{FB01} for collisions in galactic centers].
One of the possible outcomes is that the two colliding stars
merge to form a new star, which is suggested to be one of the
mechanisms to form blue stragglers [for other blue straggler formation
mechanisms, see \citet{L89}, \citet{S93} and \citet{B95} etc.].
Here, we simply call the merger products ``blue stragglers'' (BSs),
whether or not they really correspond to the BSs found in
observations.
The BS is brighter and bluer than its parent stars
for two reasons: it may have a larger mass,
or the stellar envelope may have a higher helium content if its parent
stars have evolved significantly before the collision and the central
helium is mixed to the surface of the BS during the collision
(e.g., \citealt{Sills97,Sills01}).
The answers to the questions whether collisions of main-sequence stars can
form BSs, how much mass is released from stars during stellar coalescence,
and how much helium is mixed during the collisions, depend on
the masses and radii of the colliding stars, their relative velocity and their
impact parameter etc.
Hence, these factors also determine what the luminosity properties of BSs
formed by stellar collisions will be.
Collisions of two stars with low relative velocity (e.g., $\la v\ij$ in
eq.~\ref{eq:Sigmaij}) are more likely to lead to stellar coalescence and form
BSs; and those with high relative velocity (e.g., at $r\la 10^{-3}\pc$ in
galaxies with $M_\bullet\simeq 10^6\msun$ or $r\la 1\pc$ in galaxies
with $M_\bullet\simeq 10^9\msun$ where the velocity dispersion
is significantly higher than $v\escsun$) are more likely to destroy stars,
or the stars may pass through each other and both survive the collision
(see \citealt{FB01,LRS93,BH87,BH92}).
Below we will consider several extreme cases of collision products
to study the effects of collisions on color gradients. 
None of the extreme cases are realistic, but they will
simplify the complexity of the collision outcome and provide
useful limits to the effects caused by stellar collisions.

For the mass of the collision product, we consider two extreme cases:
one is that collisions always destroy both of the colliding stars;
the other is that collisions always produce BSs
and the mass of the BS is the sum of the masses of
its parent stars.

We next consider the helium abundance of BSs.
A star which has been on the main sequence for time $\tau$ has created
a mass of helium equal to $\tau\langle L\rangle/(\epsilon c^2)$,
where $\langle L\rangle$ is the mean luminosity of the star during its
main-sequence phase so far, and $\epsilon=0.007$ is the efficiency of
hydrogen fusion.
We define $Y_0$ and $Z_0$ as the initial helium abundance and metallicity
of the stars.
If a BS is formed by the collision of two main-sequence stars at time $t$
and the chemical abundance is fully mixed
(which is the extreme case of mixing), the resulting helium abundance
$Y_{ij}$ of the BS will be
(see eq.~2 in \citealt{BP95}):
\be
Y_{ij}=Y_0+t_{10}\frac{0.1\msun}{m_i+m_j}\frac{\langle L_i\rangle+\langle L_j\rangle}{L_\odot}(1-Y_0),
\label{eq:Yij}
\ee
where $t_{10}$ is the time in units of $10^{10}\yr$,
and $m_i$ and $m_j$ are the masses of the two colliding stars.
We will consider two extreme cases for the helium
abundance of BSs:
one is that BSs have the initial helium abundance of their
parent stars $Y_0$;
and the other is that helium is completely mixed during collisions
and BSs have the highest possible helium abundance
$Y_{ij}=0.1 t_{10}+Y_0(1-0.1 t_{10})$ (eq.~\ref{eq:Yij} for collisions of
two main-sequence stars with solar mass).

According to the above treatment,
the luminosity of the collision product in equation (\ref{eq:Icollij}) or
(\ref{eq:Icollija}) is bounded by the following three extreme cases:
(i) $L_{ij,C}(t,t')=0$,
(ii)
\be
L_{ij,C}(t,t')=L_C(m_i+m_j,t-t')
\label{eq:Lij2}
\ee
with helium abundance $Y_0$ and metallicity $Z_0$, or
(iii) $L_{ij,C}(t,t')=L_C(m_i+m_j,t-t')$ with helium abundance
$Y_{ij}=0.1 t_{10}+Y_0(1-0.1 t_{10})$
and metallicity $Z_0$.
Note that the merged BSs are assumed to be equilibrium main-sequence stars
after collisions, because the BS should return to
the main sequence on the short Kelvin-Helmholtz timescale
(e.g., $\sim 10^7\yr$ for stars with solar mass).
In addition, a complete resetting of the nuclear reaction on the main
sequence assumed in equation (\ref{eq:Lij2}) (see the term ``$t-t'$'')
is unlikely, especially for merged BSs with little helium mixing, which
have a shorter main-sequence lifetime (\citealt{Sills97,Sills01}); 
and this assumption in equation (\ref{eq:Lij2}) will give an upper limit
to the effect of stellar collisions on color profiles.

\subsection{Color gradients in M32}\label{subsec:M32}

\noindent
In this subsection, we use the dwarf elliptical galaxy M32 in the Local Group
as an example to study the
color gradients caused by stellar collisions in galactic centers.
M32 (NGC 221) is useful for studying stellar collisions because
(i) it is the closest elliptical galaxy,
(ii) it is compact and its density is high,
(iii) it has no visible features of dust or other anomalies (see details
in \S~\ref{subsubsec:M32obs}).
M32 has the a shorter stellar collision timescale at {\it HST} resolution
than any other nearby galaxy (see Fig.~\ref{fig:tcoll}b),
and thus we expect that M32 will have the most easily detectable color
gradients caused by stellar collisions (if any).

\subsubsection{Observations}\label{subsubsec:M32obs}

\noindent
Both photometry (in the optical/infrared bands) and spectroscopy suggest
that the center of M32 has a relatively homogeneous
stellar population without strong gradients in age or metallicity.
No evidence of features such as an inner disk, dust, or any other structure
is visible in the {\it HST} WFPC2 optical images and the 
Near Infrared Camera and Multi-Object Spectrometer (NICMOS) images
\citep{Lauer98,COR01}.
A Nuker-law fit (eq.~\ref{eq:nukerlaw}) to the $V$-band surface
brightness profile at $r<1\arcsec$ gives $\alpha=1.39\pm0.82$,
$\beta=1.47\pm0.16$, $\gamma=0.46\pm0.14$, $I\b$ corresponding to
$\mu\b=12.91\pm0.31$, and $r\b=0.47\arcsec\pm0.15\arcsec$ \citep{Lauer98}.
The central cusp of the surface brightness profile of M32 continues to
rise into the {\it HST} resolution limit;
thus there is only a lower limit on the central density
$\rho_0>10^7$ $\msun\pc^{-3}$ \citep{Lauer98}.
The BH mass of M32 is about $(2.5\pm0.5)\times 10^6\msun$ 
and the stellar $I$-band mass-to-light ratio (determined from dynamical models)
is $M/L_I=(1.85\pm0.15)\msun/L_{\odot,I}$ \citep{Verolme02}.
The fit to the M32 color profiles for $0.1\arcsec<r<10\arcsec$ gives an
essentially flat color index as a function of projected radius $R$
\citep{Lauer98}:
\be
\mu\obs{_{,V}}-\mu\obs{_{,I}}=(1.237\pm0.002)-(0.009\pm0.005)\log(R/1\arcsec). 
\label{eq:VI}
\ee
and
\be
\mu\obs{_{,U}}-\mu\obs{_{,V}}=(1.216\pm0.004)-(0.023\pm0.008)\log(R/1\arcsec),
\label{eq:UV}
\ee
where $\mu\obs$ is the observational surface brightness in mag\,arcsec$^{-2}$.
The absorption line indexes in the optical bands also show no radial
gradients from $1\arcsec$ to $\sim10\arcsec$ \citep{del01}.
{\it HST} NICMOS imaging of the core of M32 shows that
the near-infrared surface brightness profiles can be fitted by the same form
as that
fitted to the optical images \citep{COR01} and no strong gradients are found
in the infrared color profiles \citep{P93,COR01}.
The spatial distribution of the brightest stars in the infrared also
follows the optical light profile outside $2\arcsec$ of the nucleus of M32
\citep{Davidge00}.
The absence of gradients in both colors and absorption line indexes indicates
that the stars are coeval or that the stars have been mixed considerably
in the inner region of M32.
By fitting the mean values of the spectral indexes and colors for the
inner regions of M32 to the stellar population synthesis model in
\citet{Vazdekis96},
\citet{del01} find that the population in the core of M32 can be modeled
as a single stellar population of an intermediate age ($\simeq 4\Gyr$)
and solar abundance.
Note that the IMFs used in the model of \citet{Vazdekis96} are not exactly
the same as equations (\ref{eq:SpIMF}) and (\ref{eq:multiIMF})
(e.g., they have different lower and upper limits of the stellar mass range
etc.).
In this paper, we will focus on the results obtained by using the Salpeter
IMFs in equation (\ref{eq:SpIMF}), but
our conclusions will not be significantly affected if we use the IMFs
in \citet{Vazdekis96} or equation (\ref{eq:multiIMF}).
We will study whether collisions between main-sequence stars can
produce a color gradient in M32 in visible bands.

\subsubsection{Expected color gradients caused by collisions
between main-sequence stars}
\label{sec:colorgradM32} \label{sec:collrateM32}

\noindent
Assume that the center of M32 is composed of a single stellar population
of an intermediate age ($\simeq 4\Gyr$) with a Salpeter IMF
(eq.~\ref{eq:SpIMF}) and solar abundance
(see \S~\ref{subsubsec:M32obs}),
and that the stellar population evolves as described in \S~\ref{subsec:gendis}.
In such a system, the turn-off stars have a mass $m\tf\simeq 1.3\msun$
(set $t_{\rm MS}=4\Gyr$ in eq.~\ref{eq:tms}).

In \S~\ref{sec:BS}, we have described three extreme cases 
of the luminosity of collision products $L_{ij,C}(t',t)$.
Using equations (\ref{eq:I0}), (\ref{eq:colorgrd0})--(\ref{eq:Deltamucoll}),
and the luminosity of main-sequence
stars $L_C(m,t)$ obtained from the Padova stellar evolutionary tracks
\citep{Girardi00},
we do the numerical calculation for cases (i) and (ii) and find that
stellar collisions cannot cause observable color gradients in color
indexes $\mu_U-\mu_V$ and $\mu_V-\mu_I$ in the center of M32
(i.e., $\Delta\mu\coll{_{,U-V}}<0.02\mag$ and
$\Delta\mu\coll{_{,V-I}}<0.02\mag$ at $R>0.1\arcsec$,
see Fig.~\ref{fig:colgrd} below).
In case (iii), the evolutionary tracks of stars with the required
chemical abundance is not provided in \citet{Girardi00} or any other
published literature that we are aware of.
However, we shall argue that the difference in the chemical abundance in case
(iii) will not significantly affect our conclusions obtained in cases (ii).

The above numerical results (i.e., no observable color gradients are expected to be
caused by collisions between main-sequence stars in M32)
may be understood in the following semi-analytic way.
We consider the case in which stellar collisions always lead to
stellar coalescence, i.e., case (ii).
Using equation (\ref{eq:Lij2}), we have
$\int_0^{T\age}L_{ij,C}(T\age,t')~\d t'=
\int_0^{T\age}L_C(m_i+m_j,T\age-t')~d t'=\int_0^{T\age}L_C(m_i+m_j,t)~d t$,
and the surface brightness contributed by the BSs which are formed by
collisions of two stars with total mass higher than $m\ij$ is given by
(cf., eq.~\ref{eq:Icollija}):
\begin{eqnarray}
& & I\BS\collmij{_{,C}}(R,T\age) \nonumber \\
& = & {1\over2}\int^{m\tf}_{\max(m\ij-m\tf,0)}\d m_i\int^{m\tf}_{\max(m\ij-m_i,0
)}\d m_j\int\d a_i\int\d a_j\int_{-\infty}^\infty \d z~{\cal R}_{ij}(r,T\age) \nonumber \\
& & \times \int_0^{T\age} L_C(m_i+m_j,t)~\d t
\label{eq:IcollijaBS} \\
&=& J\BS\collmij{_{,C}}(R,T\age)Rf\coll(R)
\label{eq:IcollijbBS}
\end{eqnarray}
where $J\BS\collmij{_{,C}}(r,T\age)$ is the luminosity density 
contributed by the BSs which are formed by collisions
of two stars with total mass higher than $m\ij$ at galactic radius $r$:
\begin{eqnarray}
& & J\BS\collmij{_{,C}}(r,T\age) \nonumber \\
& \equiv & {1\over2}\int^{m\tf}_{\max(m\ij-m\tf,0)}\d m_i\int^{m\tf}_{\max(m\ij-m_i,0)}\d m_j\int\d a_i\int\d a_j~{\cal R}_{ij}(r,T\age) \nonumber \\
& &
\times \int_0^{T\age} L_C(m_i+m_j,t')~\d t'.
\label{eq:JcollijaBS}
\end{eqnarray}
Here $f\coll(R)$ is a factor reflecting the projection effect of the
surface brightness.
According to equations (\ref{eq:Sigmaij})--(\ref{eq:calRij}) and
(\ref{eq:IcollijaBS})--(\ref{eq:JcollijaBS}), we approximately have
\be
f\coll(R)\simeq\frac{\int_{-\infty}^\infty n^2(r,T\age)\sigma(r)[1+v\escsun^2/6\sigma^2(r)]~\d z}{n^2(R,T\age)\sigma(R)[1+v\escsun^2/6\sigma^2(R)]R},
\label{eq:fBS}
\ee
where $v\ij\simeq v\escsun$, $\langle v\rel\rangle\propto \sigma(r)$
and $\langle v\rel\rangle\langle v\rel^{-1}\rangle^{-1}
\simeq \langle v\rel^2\rangle=6\sigma^2(r)$
[$f\coll(R)$ is not sensitive to the factor ``6'' in front of $\sigma^2(r)$]
are assumed (see eq.~\ref{eq:Sigmaij}).
Thus, as in the derivation in equations
(\ref{eq:colorgrd0coll})--(\ref{eq:colorgrdcoll}),
we may obtain the change of the color index caused by BSs given by:
\begin{eqnarray}
\Delta\mu\BS\coll{_{,C_1-C_2}}(R)
\simeq \frac{f\coll(R)}{f_0(R)}\int\d \ln m\ij~\frac{-\d J\BS\collmij{_{,C_1}}(R)/\d \ln m\ij}{J_{0,C_1}(R)}
\Delta\mu\BS\coll{_{,m\ij,C_1-C_2}}(R) \label{eq:colorgrdcollapproBS}
\label{eq:deltamuBS}
\end{eqnarray}
where
\begin{eqnarray}
\frac{\d J\BS\collmij{_{,C_1}}(R)}{\d \ln m\ij} & = &
-{m\ij \over2}\left.\int^{m\tf}_{\max(m\ij-m\tf,0)}\d m_i
\int\d a_i\int\d a_j~{\cal R}_{ij}(r,T\age)\right|_{m_j=m\ij-m_i}  \nonumber \\
& & \times \int_0^{T\age} L_C(m\ij,t)~\d t,
\label{eq:dJcollijaBS}
\end{eqnarray}
\be
\Delta\mu\BS\coll{_{,m\ij,C_1-C_2}}(R)=
[\mu\BS\coll{_{,m\ij,C_1}}(R)-\mu\BS\coll{_{,m\ij,C_2}}(R)]-[\mu_{0,C_1}(R)-\mu_{0,C_2}(R)],
\label{eq:deltamuBSa}
\ee
\begin{eqnarray}
& & \mu\BS\coll{_{,m\ij,C_1}}(R)-\mu\BS\coll{_{,m\ij,C_2}}(R) \nonumber \\
& = &
-2.5\log\left[\frac{\d J\BS\collmij{_{,C_1}}(R)/\d \ln m\ij}{\d J\BS\collmij{_{,C_2}}(R)/\d \ln m\ij}\right]+(M_{\odot,C_1}-M_{\odot,C_2}) \nonumber \\
& = &
-2.5\log\left[\frac{\int_0^{T\age} L_{C_1}(m\ij,t)~\d t}
{\int_0^{T\age} L_{C_2}(m\ij,t)~\d t}\right]+(M_{\odot,C_1}-M_{\odot,C_2}).
\label{eq:colorgrdcollBS}
\end{eqnarray}
Equation (\ref{eq:colorgrdcollapproBS}) is valid for
$[f\coll(R)J\BS\coll{_{,>0,C_1}}(R)]/[f_0(R)J_{0,C_1}(R)]\ll 1$,
$[f\coll(R)J\BS\coll{_{,>0,C_2}}(R)]/$ $[f_0(R)J_{0,C_2}(R)]\ll 1$ 
and $|\Delta\mu\BS\coll{_{,m\ij,C_1-C_2}}(R)|\ll 2.5$.
Equation (\ref{eq:deltamuBS}) shows that
for the collisions of two stars with total mass in the range
$\ln m\ij\rightarrow \ln m\ij+\d \ln m\ij$,
whether the effect of BSs on
the color gradient is significant at projected radius $R$
is controlled by three factors: 
(i) the ratio of the luminosity density
$\d \ln m\ij(\d J\collmij\BS{_{,C_1}}/\d \ln m\ij)/J_{0,C_1}$
at galactic radius $r=R$;
(ii) the difference between the color of BSs and
the original color of galactic centers,
$\Delta\mu\BS\coll{_{,m\ij,C_1-C_2}}$;
and (iii) the projection effect $f\coll/f_0$
(see eqs.~\ref{eq:f0} and \ref{eq:fBS}).

\begin{figure}
\begin{center}
\includegraphics[width=0.8\textwidth,angle=0]{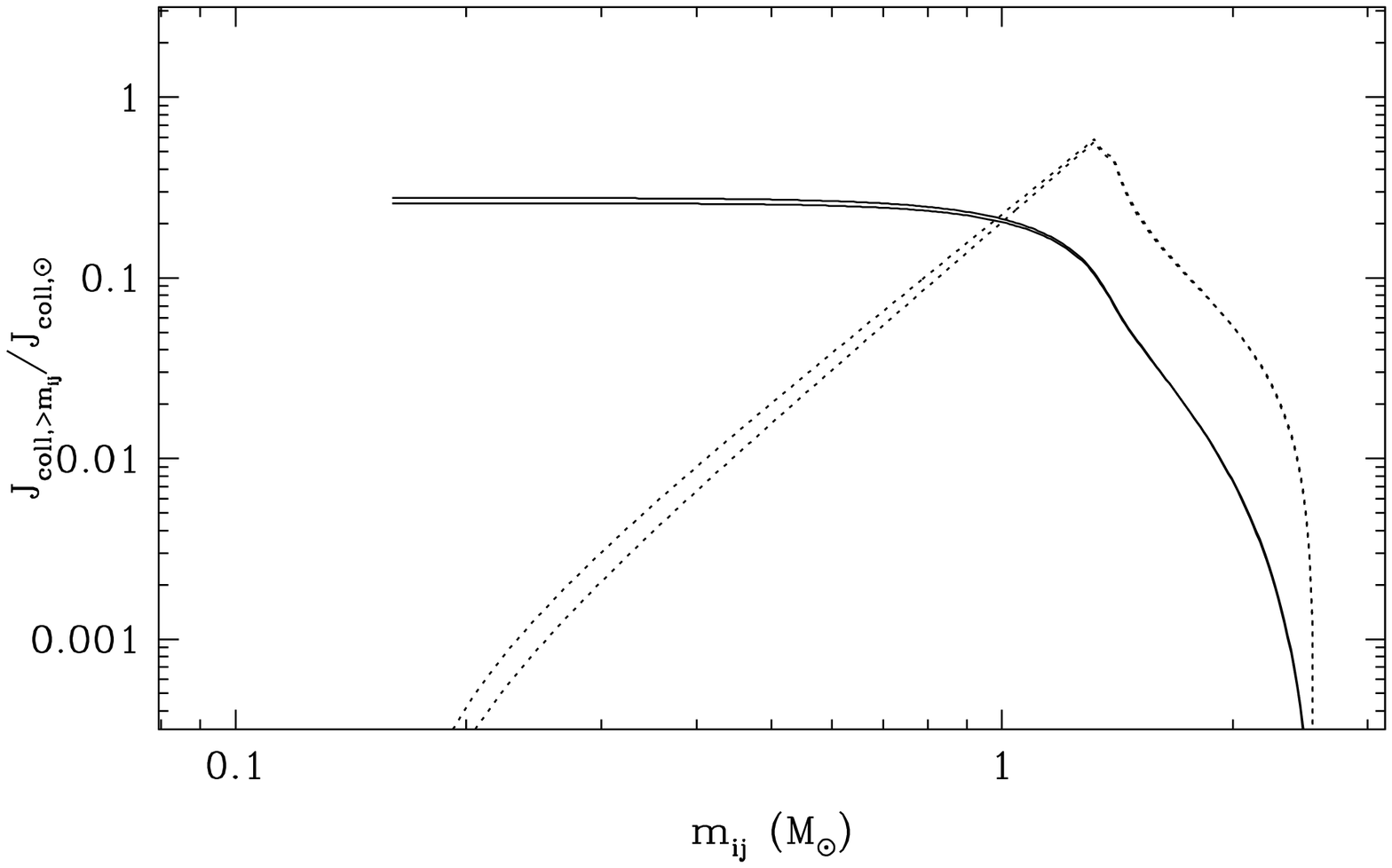}
\caption{The ratio of the luminosity density contributed by BSs formed by
collisions between main-sequence stars,
$J\BS\collmij{_{,C}}/J\BS\collsun{_{,C}}$,
as a function of stellar mass $m\ij$,
in M32 (solid lines for $\sigma(r)\ll v\escsun$ or $\sigma(r)\gg v\escsun$,
cf., eq.~\ref{eq:lumicollij}).
The $J\BS\collmij{_{,C}}$ represents the luminosity density contributed by
BSs which are formed by collisions of two main-sequence stars with total
mass higher than mass $m\ij$, and M32 is assumed to have a Salpeter IMF
(eq.~\ref{eq:SpIMF}) and age $T\age=4\times10^9\yr$
(see \S~\ref{subsubsec:M32obs}).
The $J\BS\collsun{_{,C}}$ represents the luminosity density contributed by
BSs formed by collisions by assuming that the center of M32 is composed of
identical stars with solar mass and radius.
The dotted lines represent
$|\d J\BS\collmij{_{,C}}/\d \ln m\ij|/J\BS\collsun{_{,C}}$;
and their peaks are located at $m\ij\simeq m\tf$, which shows
that most of the surface brightness contributed by BSs comes from
collisions between two stars with total mass around $m\tf$.
}
\label{fig:lumicollij}
\end{center}
\end{figure}

Below, we will give a quantitative estimate to the three factors in 
equation (\ref{eq:deltamuBS}).
If the center of M32 is composed of identical stars with solar mass and radius,
we may use the surface brightness profile of M32 and its mass-to-light
ratio $\Upsilon$ determined from dynamical models to get its stellar mass
density or stellar number density $n_\odot(r)$.
In such a system, we define the luminosity density contributed by BSs formed
by collisions between solar-type stars as $J\collsun\BS{_{,C}}(r)$; and
according to equation (\ref{eq:Icollija}), we have
\begin{eqnarray}
\frac{J\collsun\BS{_{,C}}(r)}{J_{0,C}(r)}
& \simeq & \frac{[n_\odot(r)/2 t\collsun(r)]\int_0^{T\age} L_C(2\msun)~\d t}
{n_\odot(r)L_C(\msun)/\Upsilon_C}
\simeq \frac{\Upsilon_C t_{\rm MS}(1\msun)}{t\collsun(r)} \nonumber \\
& \simeq & 0.2
\frac{10^{11}\yr}{t\collsun(r=0.1\arcsec)}
\frac{t\collsun(r=0.1\arcsec)}{t\collsun(r)}\frac{\Upsilon_C}{2}, 
\label{eq:idenIcolli}
\end{eqnarray}
where $\int_0^{T\age} L_C(2\msun)~\d t\simeq L_C(1\msun)t_{\rm MS}(1\msun)$ 
is assumed (i.e. the time-integrated luminosity of main-sequence stars is
assumed to have $\int_0^{T\age} L_C(m,t)~\d t\propto m$ for $m\ge m\tf$),
and $t_{\rm MS}(1\msun)=10^{10}\yr$.
If the stars have a distribution of masses and radii,
according to equation (\ref{eq:JcollijaBS}),
we have the ratio of the luminosity density contributed by BSs as follows
(similarly as eqs.~\ref{eq:tcollmisun}, \ref{eq:dotrhocollisun} and
\ref{eq:dotrhocollijsun}):
\begin{eqnarray}
& & \frac{J\BS\collmij{_{,C}}(r)}{J\BS\collsun{_{,C}}(r)} \nonumber \\
& \simeq & \left[\int_0^\infty\d m_i~(m_i/\msun)\Xi(m_i,T\age)\right]^{-2}
\nonumber \\
& & \times 
\int^{m\tf}_{\max(m\ij-m\tf,0)}\d m_i\int^{m\tf}_{\max(m\ij-m_i,0)}\d m_j~\Xi(m_i,0) \Xi(m_j,0)
\frac{\int_0^{T\age}L_C(m_i+m_j,t)~\d t}{\int_0^{T\age}L_C(2\msun,t)~\d t}
\nonumber \\
& & \times \cases{[(a_i+a_j)/(2\rsun)]^2 & if $\sigma(r)\gg v\escsun$ \cr
     [(a_i+a_j)/(2\rsun)][(m_i+m_j)/(2\msun)] & if $\sigma(r)\ll v\escsun$ \cr
}.
\label{eq:lumicollij}
\end{eqnarray}
We assume the following relation for the luminosity of main-sequence stars:
\be
\int_0^{T\age} L_C(m,t)~\d t\propto
\cases{ m & if $m\ge m\tf$ \cr
        m^{\eta_L} & if $m<m\tf$ \cr
}.
\label{eq:LC}
\ee
(i.e., the luminosity of main-sequence stars is assumed to be
$L_C(m,t)\propto m^{\eta_L}$; $\eta_L=3.5$).
Using equations (\ref{eq:lumicollij}) and (\ref{eq:LC}),
we show in Figure~\ref{fig:lumicollij} the ratio of the luminosity density
$J\BS\collmij{_{,C}}/J\BS\collsun{_{,C}}$ (solid lines)
and its derivative
$|\d J\BS\collmij{_{,C}}/\d \ln m\ij|$ $/J\BS\collsun{_{,C}}$
(dotted lines)
as a function of stellar mass $m\ij$.
We show the results for both of the two cases $\sigma(r)\gg v\escsun$
and $\sigma(r)\ll v\escsun$ in equation (\ref{eq:lumicollij}), 
and we have
\be
\frac{J\BS\coll{_{,>0,C}}(r)}{J\BS\collsun{_{,C}}(r)}\simeq 0.3.
\label{eq:Splumicollmi}
\ee
The peak of the derivative
$|\d J\BS\collmij{_{,C}}/\d \ln m\ij|/J\BS\collsun{_{,C}}$
is located at $m\ij\simeq m\tf$, which represents that 
most of the luminosity contributed by BSs comes from
collisions between two main-sequence stars with total mass $m\ij\simeq m\tf$.

\begin{figure}
\begin{center}
\includegraphics[width=0.8\textwidth,angle=0]{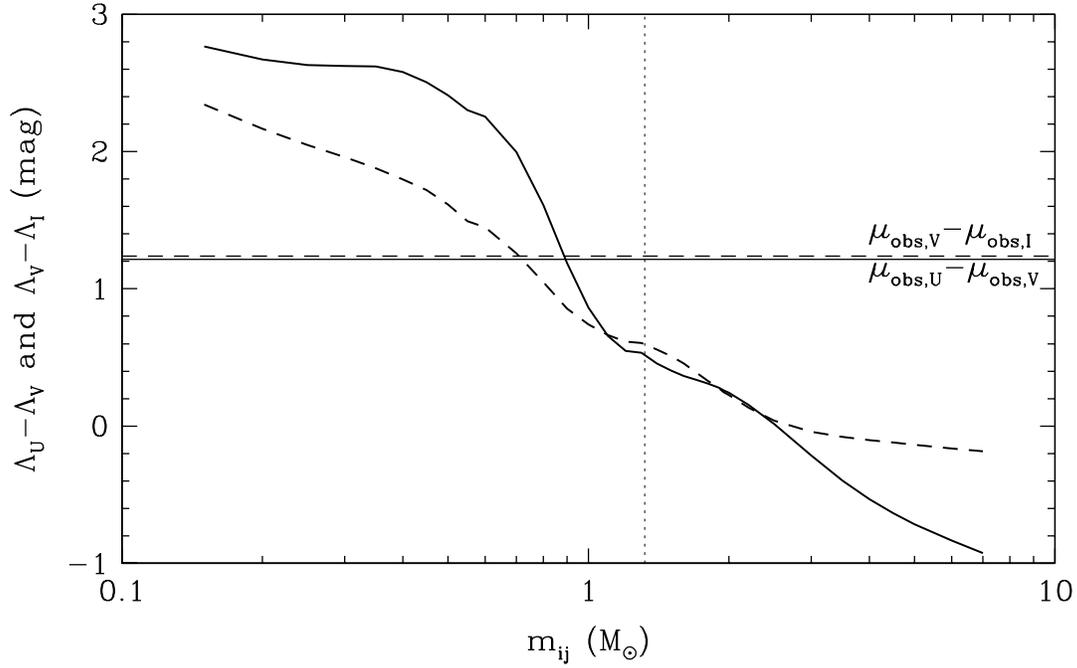}
\caption{The color indexes $\Lambda_U-\Lambda_V$ (thick solid line) and 
$\Lambda_V-\Lambda_I$ (thick dashed line) of the time-integrated luminosity
($\int_0^{T\age} L_C~\d t$) of stars during their main-sequence phases,
as a function of stellar mass $m\ij$ (eq.~\ref{eq:Ldt}).
The results are based on the stellar evolutionary tracks of \citet{Girardi00}.
The stars have solar abundance.
The horizontal lines give the observed color indexes, 
$\mu\obs{_{,U}}-\mu\obs{_{,V}}$ (thin solid line) and
$\mu\obs{_{,V}}-\mu\obs{_{,I}}$ (thin dashed line)
of the surface brightness of M32 at $R=1\arcsec$
(eqs.~\ref{eq:VI} and \ref{eq:UV}).
The vertical dotted line represents stellar mass $m\ij=1.3\msun$, which
is the turn-off mass
at age $T\age\simeq 4\Gyr$ (see eq.~\ref{eq:tms}).
}
\label{fig:colorind}
\end{center}
\end{figure}

To obtain the difference between the color of BSs and the original colors
of the center in M32 in the absence of stellar collisions
(i.e., $\Delta\mu\coll\BS{_{,m\ij,C_1-C_2}}$ in eq.~\ref{eq:deltamuBSa}),
we use the Padova stellar evolutionary track \citep{Girardi00}
to obtain the color indexes of time-integrated luminosity
$\int_0^{T\age} L_C(m\ij,t)~\d t$ of stars
during their main-sequence phases as a function of stellar mass $m\ij$, 
$\Lambda_{C_1}(m\ij)-\Lambda_{C_2}(m\ij)$, defined by
(cf., eq.~\ref{eq:colorgrdcollBS}):
\begin{eqnarray}
\Lambda_{C_1}(m\ij)-\Lambda_{C_2}(m\ij)
\equiv -2.5\log\left[\frac{\int_0^{T\age} L_{C_1}(m\ij,t)~\d t}
{\int_0^{T\age} L_{C_2}(m\ij,t)~\d t}\right]+(M_{\odot,C_1}-M_{\odot,C_2})
\label{eq:Ldt}
\end{eqnarray}
(in the integration $\int_0^{T\age} L_C(m\ij,t)~\d t$, even if we include
the luminosity during the post-main-sequence phase, our conclusion will
not be affected).
Combining equations (\ref{eq:colorgrdcollBS}) and (\ref{eq:Ldt}), we have
\be
\mu\BS\coll{_{,m\ij,C_1}}-\mu\BS\coll{_{,m\ij,C_2}}=\Lambda_{C_1}(m\ij)-\Lambda_{C_2}(m\ij).
\ee
The $\Lambda_U(m\ij)-\Lambda_V(m\ij)$ and $\Lambda_V(m\ij)-\Lambda_I(m\ij)$
for stars with solar abundance are shown as a function of stellar mass
$m\ij$ in Figure~\ref{fig:colorind}.
The horizontal lines in Figure~\ref{fig:colorind} give the observed
color indexes of $\mu\obs{_{,U}}-\mu\obs{_{,V}}$ (thin solid line) and
$\mu\obs{_{,V}}-\mu\obs{_{,I}}$ (thin dashed line)
in the center of M32 at $R=1\arcsec$ (eqs.~\ref{eq:VI} and \ref{eq:UV}).
As seen from Figure~\ref{fig:colorind},
for color index $\Lambda_U(m\ij)-\Lambda_V(m\ij)$ (thick solid line),
the stars with mass $m\ij\ga 0.9\msun$ are bluer
than the center of M32;
and for color index $\Lambda_V(m\ij)-\Lambda_I(m\ij)$ (thick dashed line),
the stars with mass $m\ij\ga 0.7\msun$ are bluer
than the galactic center of M32.
In Figure~\ref{fig:colorind},
at $m\ij\simeq m\tf$, where the peak of 
$|\d I\BS\collmij{_{,C}}/\d \ln m\ij|/I\collsun{_{,C}}$
is located, we have 
\be
\Delta\mu\coll\BS{_{,m\ij,U-V}}=
[\Lambda_U(m\ij)-\Lambda_V(m\ij)]-[\mu\obs{_{,U}}-\mu\obs{_{,V}}]\simeq -0.7
\label{eq:DeltamucollUV}
\ee
and
\be
\Delta\mu\coll\BS{_{,m\ij,V-I}}
=[\Lambda_V(m\ij)-\Lambda_I(m\ij)]-[\mu\obs{_{,V}}-\mu\obs{_{,I}}]\simeq -0.6.
\label{eq:DeltamucollVI}
\ee

\begin{figure}
\begin{center}
\includegraphics[width=0.8\textwidth,angle=0]{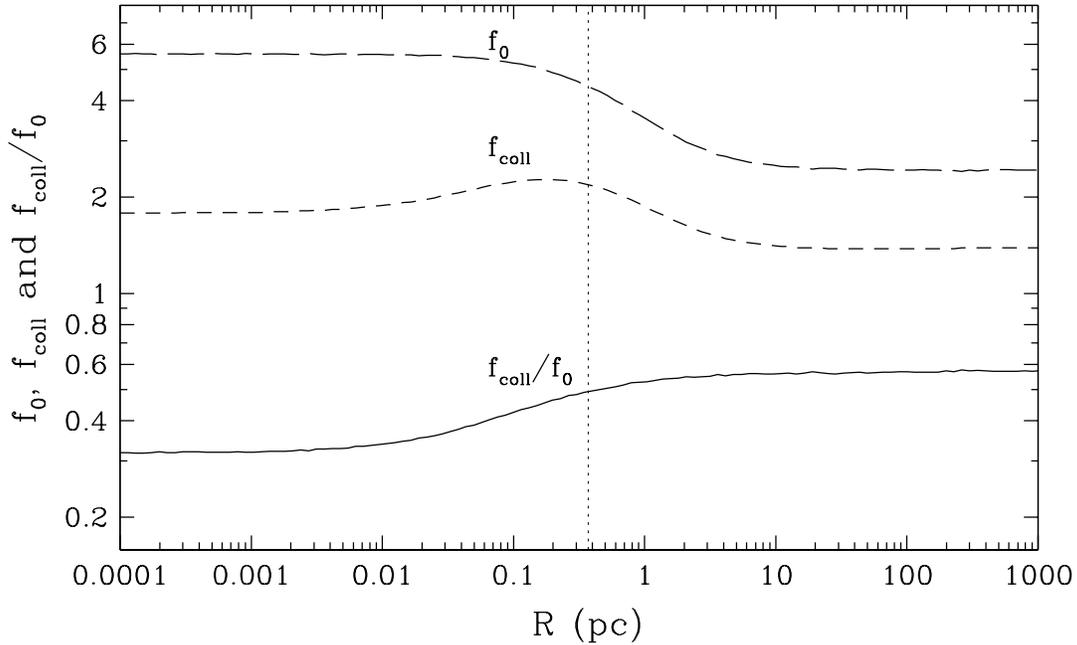}
\caption{The projection factors $f_0(R)$ (long dashed line),
$f\coll(R)$ (short dashed line) and their ratios 
$f\coll(R)/f_0(R)$ (solid line) as a function of projected radius
$R$ in M32 (see eqs.~\ref{eq:f0} and \ref{eq:fBS}).
The vertical dotted line represents $R=0.1\arcsec$.
The results in the region not resolved by the {\it HST}
($2R\la 0.1\arcsec$) are obtained by
extrapolating the Nuker-law-fitted surface brightness profile inwards.
}
\label{fig:fproj}
\end{center}
\end{figure}

Using equations (\ref{eq:f0}) and (\ref{eq:fBS}),
we obtain the projection factors $f_0(R)$ and $f\coll(R)$ of M32 and
their ratio $f\coll(R)/f_0(R)$ as a function of projected radius
$R$, which are shown in Figure~\ref{fig:fproj}.
The results in the region not resolved by the {\it HST} ($2R<0.1\arcsec$)
are obtained by extrapolating the Nuker-law fitted
surface brightness profile inwards.
Figure~\ref{fig:fproj} shows that $f\coll(R=0.1\arcsec)/f_0(R=0.1\arcsec)$
weakly depends on $R$ and
\be
\frac{f\coll(R=0.1\arcsec)}{f_0(R=0.1\arcsec)}\simeq 0.5.
\label{eq:fratio}
\ee

Using equation (\ref{eq:deltamuBS}),
we have the following relation for the change of the color index caused by
stellar collisions:
\be
|\Delta\mu\BS\coll{_{,C_1-C_2}}(R)|<
\frac{f\coll(R)}{f_0(R)}
\frac{J\BS\coll{_{>0,C_1}}(R)}{J_{0,C_1}(R)}
|\Delta\mu\BS\coll{_{,m\tf,C_1-C_2}}|,
\label{eq:deltafinal}
\ee
where we use the value of $\Delta\mu\BS\coll{_{,m\ij,C_1-C_2}}$,
at $m\ij\simeq m\tf$
since the peak of $|\d I\BS\collmij/\d \ln m\ij|/I_0$ is located at
$m\ij\simeq m\tf$
(note that $|\d I\BS\collmij/\d \ln m\ij|/I_0$
decrease sharply from $m\ij\simeq m\tf$ to both
high-mass and low-mass ends in Figure~\ref{fig:lumicollij} and 
$\Lambda_{C_1}(m\ij)-\Lambda_{C_2}(m\ij)$ change mildly in the whole
mass range in Figure~\ref{fig:colorind}).
Using equations (\ref{eq:idenIcolli}), (\ref{eq:Splumicollmi}),
and (\ref{eq:DeltamucollUV})--(\ref{eq:fratio}) in equation
(\ref{eq:deltafinal}),
we have
\begin{eqnarray}
0>\left.\cases{ \Delta\mu\BS\coll{_{,V-I}}(R) \cr
          \Delta\mu\BS\coll{_{,U-V}}(R) \cr
        }\right\}>-0.03\mag \left[\frac{t\collsun(R)}{t\collsun(R=0.1\arcsec)}\right],
\label{eq:colorgrdpred}
\end{eqnarray}
where the collision timescale $t\collsun(R=0.1\arcsec)\simeq 10^{11}\yr$
(see Fig.~\ref{fig:tcoll}) for M32 is used.
Inequality (\ref{eq:colorgrdpred}) gives an upper limit of the blueing
caused by collisions between main-sequence stars,
which shows that it is almost impossible for stellar collisions
to cause observable blueing at $R\ga 0.1\arcsec$ in the center of M32.
But inequality (\ref{eq:colorgrdpred}) does not exclude
the possibility that stellar collisions can cause a color difference
$>0.02\mag$ in the much inner region at $R<0.1\arcsec$.
Note that inequality (\ref{eq:colorgrdpred}) is invalid 
in the region [e.g., $R\la 0.01\arcsec$ with $t\coll(R)\la T\age$]
where the approximation in equations (\ref{eq:colorgrdcollappro}) and
(\ref{eq:colorgrdcollapproBS}) is invalid.

In the assumed extreme case (iii) on the luminosity of collision products,
BSs have a different Helium abundance from case (ii).
But the difference in Helium abundance will not significantly affect the
luminosity and colors of BSs or the terms $J\BS\collmij{_{,C}}(R)/J_{0,C}(R)$
and
$\Delta\mu\BS\coll{_{,m\ij,C_1-C_2}}(R)$ (cf., eq.~\ref{eq:colorgrdcollapproBS}),
and hence will not affect our conclusions for the following reasons.
In cases (ii) and (iii),
the mean molecular weights $\mu\w=(2X+0.75Y+0.5Z)^{-1}$ are about
0.62 and 0.63 (Helium abundance $Y=0.28$ and 0.31), respectively.
For main-sequence stars with same mass,
the increase in the mean molecular weights can only increase
the luminosity by about eight percent
($L\propto\mu\w^4$, see eq.~20.20 in \citealt{KW90}).
The radius of the stars ($a_*\propto \mu\w^{0.6}$, see eq.~20.21 in
\citealt{KW90}) depends weakly on the mean molecular weight.
Thus, the increase in the mean molecular weight and stellar radius will cause
the increase of the effective temperature of the star 
($T_{\rm eff}\propto L/R^2\propto\mu\w^{2.8}$)
by about five percent, which may make main-sequence stars with mass around
$m\tf$ (with effective temperatures $T_{\rm eff}\sim 6\times 10^3\K$)
bluer only by less than $0.1\mag$
in $\Lambda_U-\Lambda_V$ and $\Lambda_V-\Lambda_I$
(or $\Delta\mu\BS\coll{_{,m\tf,U-V}}$ and 
$\Delta\mu\BS\coll{_{,m\tf,V-I}}$).
Therefore, the results obtained for case (iii) (considering the change of the
luminosity and colors of BSs caused by helium mixing) should not be
significantly different from those obtained for case (ii).

Inequality (\ref{eq:colorgrdpred}) only provides the effect on color
indexes caused by BSs.
Including the effect of removing the luminosity of parent stars
from the galactic center
[i.e., ``the term $-L_C(m_i,a_i,T\age)-L_C(m_j,a_j,T\age)$''
in eq.~\ref{eq:Icollij}] will not make the galactic center become bluer
because the decrease of the luminosity caused by destruction of
main-sequence stars comes mainly from destruction of high-mass stars
(just as most of the mass density involved in collisions come from
destruction of high-mass stars shown in Fig.~\ref{fig:lumicollij}),
which are usually bluer than the center of M32
(cf., Fig.~\ref{fig:colorind}).
If collisions of main-sequence stars are assumed to always lead
to destruction of main-sequence stars and no BSs are formed [i.e. case (i)],
using the above similar analysis for the effects by BSs,
we find that there will be no observable reddening at $R>0.1\arcsec$
of M32, either.

Figure~\ref{fig:colgrd} shows our numerical results for the color indexes
$\mu_V(R)-\mu_I(R)$ and $\mu_U(R)-\mu_V(R)$ in M32, obtained by using equations
(\ref{eq:I0}), (\ref{eq:colorgrd0})--(\ref{eq:Deltamucoll}) and
considering the luminosity of collision products in case (i) and case (ii)
(dashed line and solid line) as described in \S~\ref{sec:BS}.
We use the observed color indexes (eqs.~\ref{eq:VI} and \ref{eq:UV})
as the original color index $\mu_{0,C_1}-\mu_{0,C_2}$.
Figure~\ref{fig:colgrd} shows that collisions of main-sequence stars
cannot cause observable color changes
in the visible bands at $R>0.1\arcsec$ in M32,
which is consistent with the observation that no color gradients are seen
in M32 by {\it HST};
and at even smaller radii,
stellar collisions are likely to cause the color difference
larger than $>0.02\mag$ (e.g., at most $0.08\mag$ at $R=0.02\arcsec$).
The above analysis and numerical results are only valid
if  $t\coll(R)\ga T\age$,
In the region with $t\coll(R)\la T\age$
(e.g., $R\la 0.01\arcsec$), it is possible that most of the main-sequence
stars have experienced collisions.
If we extrapolate the observed $V$-band surface brightness profile inward
and assume that the surface brightness in the region with
$t\collsun(R)\la T\age$ has the color of BSs described in equations
(\ref{eq:DeltamucollUV}) and (\ref{eq:DeltamucollVI}),
and then use the above numerical results of the surface brightness
in the region with $t\collsun(r)>T\age$,
we may obtain the averaged color index within a given galactic projected
radius $R$.
We find for case (ii) that the averaged blueing within $R<0.1\arcsec$ due to
collisions is not larger than $0.06\mag$ in $\mu_U-\mu_V$ and $0.02\mag$ in
$\mu_V-\mu_I$, and the averaged blueing within $R<0.05\arcsec$ due to
collisions is not larger than $0.16\mag$ in $\mu_U-\mu_V$ and $0.06\mag$ in
$\mu_V-\mu_I$.

In \citet{Lauer98}, stellar collisions are claimed to be capable of causing
observable blueing in the center of M32.
However, Lauer et al.\ did not account carefully for
distribution of stellar masses and radii, projection effects,
and the difference between the color of BSs formed by stellar collisions
and the original color of the center of M32.

Based on the above results for M32, we expect that stellar collisions cannot
cause any observable color gradients at $R>0.1\arcsec$ in visible bands
in the centers of M31 and other nearby galaxies.
This result may be understood as follows:
these systems have longer collision timescales than M32
(cf., Figure.~\ref{fig:tcoll}b or Table 6 in \citealt{Lauer98});
compared to the values obtained for M32,
the factor $|\Delta\mu\BS\coll{_{,m\tf,C_1-C_2}}|$
(the difference between the color of BSs with mass $m\tf$ and the original
color of galactic centers) in equation
(\ref{eq:deltafinal}) should not change significantly;
and the ratios of the projection factors $f\coll(R\ll r\b)/f_0(R\ll r\b)$
for core galaxies with small $\gamma$ may become even smaller,
and those for power-law galaxies do not significantly increase.
Multicolor {\it HST} WFPC2 images show that the center of M31
appears bluer in a $0.14\arcsec\times 0.14\arcsec$ box
than the rest of the nucleus and the surrounding bulge
($1\arcsec<R<5\arcsec$), by $\sim 0.1\mag$ in $\mu_V-\mu_I$
(see Table 2 in \citealt{Lauer98}).
Our estimate shows that the averaged blueing is not larger than
$0.03\mag$ in $\mu_V-\mu_I$ within $R<0.14\arcsec/2$ even for M32,
and hence the $\sim 0.1\mag$ blueing in M31 is unlikely to be caused by
stellar collisions.
It is therefore worthwhile to investigate other mechanisms
which may cause the blueing in the center of M31.

\begin{figure}
\begin{center}
\includegraphics[width=0.8\textwidth,angle=0]{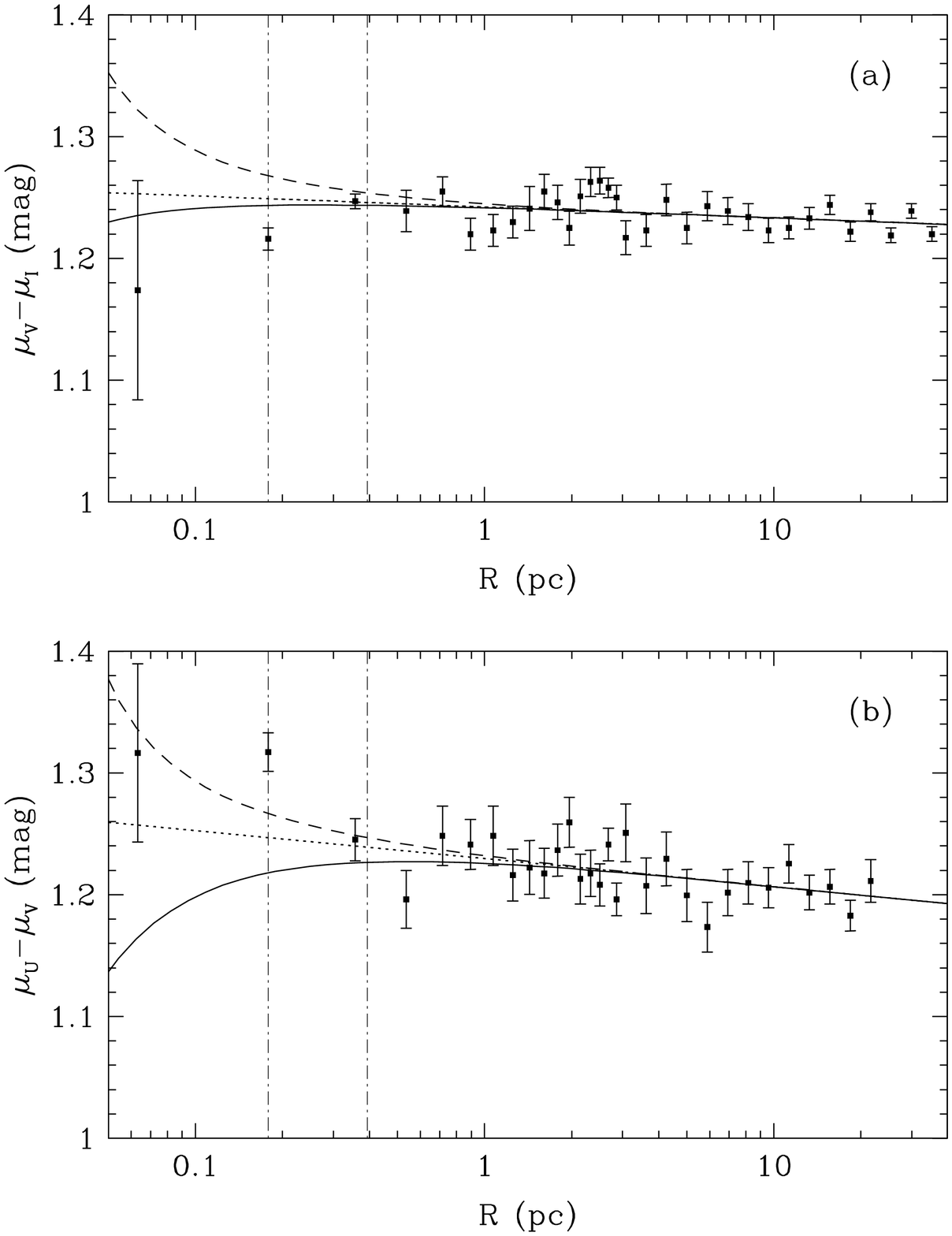}
\caption{The estimated color indexes $\mu_V-\mu_I$ and $\mu_U-\mu_V$
(eq.~\ref{eq:Icollija}) in the center of M32,
as a function of projected radius $R$.
Equations (\ref{eq:I0}) and (\ref{eq:colorgrd0})--(\ref{eq:Deltamucoll})
are used in the calculation.
The dashed lines represent the color gradients obtained by assuming
that collisions always lead to destruction of both colliding stars 
and that no BSs are formed, [case (i) of \S~\ref{sec:BS}]; 
and the solid lines show the results obtained for case (ii),
in which the colliding stars are assumed to form a BS with solar abundance.
The left vertical dot-dashed line marks the {\it HST} resolution
($R=0.0455\arcsec$) and the right vertical dot-dashed line marks
$R=0.1\arcsec$.
The solid squares are the color indexes obtained from observations
(see Fig.~18 in \citealt{Lauer98});
and the dotted lines represent a fit to the observed color index
in the region $0.1\arcsec<R<10\arcsec$ (eqs.~\ref{eq:VI} and \ref{eq:UV}),
which is used as the initial state in our calculations
(i.e., $\mu_{0,C_1}-\mu_{0,C_2}$ in eq.~\ref{eq:colorgrd0} and 
\ref{eq:Deltamucoll}).
}
\label{fig:colgrd}
\end{center}
\end{figure}

\section{Conclusions}\label{sec:discon}

\noindent
We have studied the effects of stellar collisions
in realistic galactic centers,
focusing on feeding massive BHs and color gradients.

We find that the mass involved in stellar collisions is not sufficient to
provide the present BH mass in realistic galactic centers,
especially in bright core galaxies.
Similarly, \citet{MT99} showed that stellar tidal disruption rates
cannot contribute significantly to the present BH mass in galaxies
brighter than $\sim10^9\Lsun$.
These two results together discredit two proposed mechanisms
(stellar collisions and tidal disruption of stars by central BHs,
e.g., \citealt{F78}) for the growth of massive BHs in the centers in
bright galaxies.
The raw material for BH growth must therefore come from other sources,
for example, the mass released by stellar evolution in the initial $\sim1\Gyr$
of the galaxy's lifetime or gas that sinks to the galactic centers
in a merger (see simulations on galaxy encounters in \citealt{BH96}).
For both of the above mechanisms,
massive BHs may form and grow rapidly, early in the history of the universe
(e.g., at redshift $z\sim$2--3, where galaxy encounters mostly occur
in the hierarchical galaxy formation model).
This scenario is consistent with the argument made by \citet{YT02}
that growth of high-mass BHs ($>10^8\msun$) comes
mainly from accretion during optically bright QSO phases
(which is obtained by studying the relation between the local BH mass function
and the QSO luminosity function).

We have analyzed how the color of a stellar system
is affected by collisions of main-sequence stars.
We account for the effects of collisions of main-sequence stars on color indexes
at projected radius $R$ due to three factors
(see eq.~\ref{eq:colorgrdcollapproBS}):
(i) the ratio of the change of luminosity density caused by stellar collisions
to the original luminosity density of galactic centers
at galactic radius $r=R$;
(ii) the difference between the colors of BSs and destroyed stars and
the original color of galactic centers;
and (iii) the projection effect between the surface brightness and
the luminosity density. 
We find that collisions between main-sequence stars cannot cause observable
color gradients in the visible bands at $R>0.1\arcsec$ in M32, M31 and other
nearby galactic centers.
This result is consistent with the 
lack of an observed color gradient in M32 at {\it HST} resolution.
At even smaller radii,
collisions between main-sequence stars are likely to cause the color difference
larger than $>0.02\mag$ (e.g., at most $0.08\mag$ at $R=0.02\arcsec$).
The averaged blueing due to stellar collisions in the region $R<0.1\arcsec$
of M32 should not be larger than $0.06\mag$
in $\mu_U-\mu_V$ and $0.02\mag$ in $\mu_V-\mu_I$.
The observed blueing in the center of the galaxy M31
(in a $0.14\arcsec\times 0.14\arcsec$ box; \citealt{Lauer98}) must be caused
by some mechanisms other than collisions between main-sequence stars.

This project was suggested by Scott Tremaine, my thesis advisor;
and I am deeply indebted to him for his advice.
I also thank Marc Freitag, Jeremy Goodman, Alison Sills 
and David Spergel for helpful discussions.
Support for Proposal number HST-AR-09513.01-A was
provided by NASA through a grant from the Space Telescope Science Institute,
which is operated by the Association of Universities for Research in
Astronomy, Incorporated, under NASA contract NAS5-26555. Support was also
provided by NSF grant AST-9900316.

\end{document}